\documentclass{emulateapj}

\usepackage{amsmath,amssymb}
\usepackage{subfigure,epsfig,url,bm,amsmath,cases,color,soul,longtable}

\begin{document}

\title{The Influence of Passing Field Stars on the Solar System's Dynamical Future}

\author{Nathan A. Kaib}
\affiliation{Planetary Science Institute, 1700 E. Fort Lowell, Suite 106, Tucson, AZ 85719, USA}
\author{Sean N. Raymond}
\affiliation{Laboratoire d'Astrophysique de Bordeaux, CNRS and Universit\'e de Bordeaux, All\'ee Geoffroy St. Hilaire, 33165 Pessac, France}

\begin{abstract} 

The long-term dynamical future of the Sun's planets has been simulated and statistically analyzed in great detail, but most prior work considers the solar system as completely isolated, neglecting the potential influence of field star passages. To understand the dynamical significance of field star encounters, we simulate several thousand realizations of the modern solar system in the presence of passing field stars for 5 Gyrs. We find that the impulse gradient of the strongest stellar encounter largely determines the net dynamical effect of field stars. Because the expected strength of such an encounter is uncertain by multiple orders of magnitude, the possible significance of field stars can be large. Our simulations indicate that isolated models of the solar system can underestimate the degree of our giant planets' future secular orbital changes by over an order of magnitude. In addition, our planets and Pluto are significantly less stable than previously thought. Field stars transform Pluto from a completely stable object over 5 Gyrs to one with a $\sim$5\% instability probability. Furthermore, field stars increase the odds of Mercury's instability by $\sim$50--80\%. We also find a $\sim$0.3\% chance that Mars will be lost through collision or ejection and a $\sim$0.2\% probability that Earth will be involved in a planetary collision or ejected. Compared to previously studied instabilities in isolated solar systems models, those induced by field stars are much more likely to involve the loss of multiple planets. In addition, they typically happen sooner in our solar system's future, making field star passages the most likely cause of instability for the next 4--4.5 Gyrs. 

\end{abstract}


\section{Introduction}

The continued stability of our Sun's planetary orbits over its main sequence lifetime is very likely but not guaranteed. Multiple simulations of the future dynamical evolution of the Sun's planets have shown that dynamical chaos drives a diffusion of the planets' orbits \citep[e.g.][]{lask94, lask08, zeeb17}. In the case of the outer giant planets, this diffusion is modest, and the secular eigenfrequencies governing the outer planets' eccentricity and inclination fluctuations diffuse by $\sim$1 part in 10$^{4-5}$ over 5 Gyrs \citep{hoanglask21}. In the inner solar system, the situation is far different, as the eccentricity oscillations of the inner planets do not appear regular and bounded, and their maximum eccentricities attained steadily diffuse on timescales of hundreds of Myrs \citep{lask08}. On multi-Gyr timescales there are small but non-negligible probabilities for eccentricity excursions of order the inner planets' current mean eccentricities. In the case of Mercury, this can culminate in a collision with the Sun or Venus \citep{lask94}. Suites of numerical simulations of the solar system have pegged the probability of such an event over 5 Gyrs at roughly 0.8--1\%, with the most likely timescale of occurrence at 4--5 Gyrs from now \citep{laskgast09, abbot23}. 

All of the above dynamical outcomes have been discovered through models of the solar system that assume it is a completely isolated system. In reality, the solar system receives perturbations from its local galactic environment, namely passing field stars and the Galactic tide \citep{opik32, heitre86}. Typical field star passages (as well as the Galactic tide) are dynamically inconsequential for all but the Sun's most weakly bound objects \citep{oort50, bras08}. However, the perturbative strength of stellar encounters can vary wildly depending on the star's mass, velocity, and impact parameter, and as time progresses, the probability for close and powerful encounters increases. In previous work, we showed that the perturbations from field star passages are powerful enough to accelerate the chaotic diffusion in the inner solar system beyond what is seen in isolated solar system models \citep{kaibray24}. This chaotic diffusion is known to ultimately be linked to the potential instability of Mercury's orbit \citep{lask94}. Moreover, other work has demonstrated that stellar passages powerful enough to perturb Neptune's orbital elements by $\sim$0.1\% can enhance the probability of Mercury's instability by an order of magnitude \citep{brownrein22}. 

Thus, assessments of the future dynamical evolution of the solar system could be incomplete if the effects of passing stars are not considered. It is well-established that very close stellar passages have the potential to dramatically reshape the solar system. For passages inside 100 au (of which there is a $\sim$5\% probability over 5 Gyrs), \citet{ray24} found a 2.5\% chance of destabilizing Mercury (via ejection or collision) and a 1.2\% chance of destabilizing Mars over the subsequent 20 Myrs following the passage. Similarly, \citet{laughadams00} performed a set of short stellar scattering experiments that found a 1 in 50000 chance that a stellar encounter will perturb Earth's eccentricity by 0.05 or higher. Finally, \citet{zink20} concluded that stellar encounters will likely ultimately destabilize our giant planets in the 30--40 Gyrs after the Sun becomes a white dwarf. 

While prior studies of stellar encounters have uncovered a great deal, each of them has limitations in folding their results into our broad understanding of the solar system's future dynamical evolution. For instance, an integration time of tens of Myrs or less is often employed to assess the effects of stellar encounters, even though it typically takes Gyrs for instabilities to develop in isolated solar system models \citep{laughadams00, ray24}. In addition, some works have excluded certain subsets of planets for both physical and computational considerations \citep{laughadams00, zink20}, and we know that instabilities in isolated models arise from complex interactions between multiple planets \citep[e.g.][]{lithwu11, hoang22}. Finally, it has been shown that a passage's impulse gradient ($\frac{2GM_{*}}{v_\infty b^2}$, where $M_{*}$ is stellar mass, $v_\infty$ is approach velocity, and $b$ is impact parameter) is a good predictor of a passage's effects on the planetary orbits \citep{ray24, kaibray24}, but the passages studied in prior works do not always reflect the full distribution of impulse gradients we expect from the Galactic field \citep{brownrein22}. (Note that the impulse gradient is based on a tidal approximation for the heliocentric impulse arising from a distant stellar passage \citep{rick76}.)

In this current work, we seek to more completely understand how field stars alter the potential dynamical fates of our planets and Pluto over the next 5 Gyrs. Our paper is broken up into the following structure: Section \ref{sec:meth} provides an overview of our numerical simulations and the motivation behind their structure. Section \ref{sec:res} details the results of our simulations. This includes consideration of Pluto's stability, the secular evolution of the giant planets, the overall stability of our planets' orbits, and the evolution of Earth. Finally, in Section \ref{sec:con} we summarize the conclusions of our work. 

\section{Dynamical Simulation Methods}\label{sec:meth}

To understand the influence of stellar passages on the dynamical evolution of the solar system, we perform five separate sets of simulations. Each set of simulations contains 1000 realizations of the modern day state of the solar system. The simulations contain all eight planets and Pluto on their heliocentric osculating 2000 January 1 elements specified within the JPL Horizons system. As in \citet{kaibray24}, in each individual realization, the mean anomaly of each planet (and Pluto) is shifted by a random amount that equates to a physical distance shift between $\pm2$ cm, much smaller than the actual uncertainty in planetary positions. Systems are integrated with the MERCURY Hybrid N-body algorithm with additional modifications to include passing stellar mass bodies \citep{cham99, kaib18, kaibray24}. In one simulation set, system integrations are repeated using the WHFast integrator available through the REBOUNDx N-body package \citep{reinliu12, reintam15, tam20}. All simulations are integrated until $t=5$ Gyrs (where $t=0$ is the present epoch) with an integration timestep of 1.5 days, comparable to or smaller than most other works exploring the long-term dynamical evolution of the solar system \citep{abbot23}. 

In our MERCURY simulations particles are removed from the simulation via ejection ($r>1$ pc), collision with the Sun ($r<0.005$ au), or collision with one another.  In the case of collisions, these events coincide with close encounters between massive bodies. During such encounters, the interaction portion of the Hamiltonian that involves the two encountering bodies is integrated simultaneously with the Keplerian portion using a Bulirsch-Stoer integration routine with an error tolerance of $10^{-15}$ \citep{bulsto80, cham99}. At the point of overlapping radii, the two objects are merged inelastically. 

\subsection{Stellar Passage Simulations}

In our first set of 1000 simulations (called {\it withstars}), each individual system is subjected to a unique set of field star passages as it evolves over 5 Gyrs. Passing field stars are introduced to our simulations at $d=1$ pc, and they are integrated until their heliocentric distance again exceeds 1 pc, at which point they are removed. Our stellar passage routine divides the local stellar population into 14 subpopulations, each with its own mean encounter frequency based on the population's local density and velocity distribution with respect to the Sun. For main sequence stars, we infer local densities using an empirical present-day mass function from \citet{reid02}, which is a combination of power laws: 
\begin{equation}
\psi(M) = \frac{dN}{dM} = kM^{-\alpha}.
\end{equation}

This mass function employs a power-law index of 1.3 between $0.05<M<0.7$ M$_{\odot}$, 2.8 between $0.7<M<1.1$ M$_{\odot}$, and 4.8 between $1.1<M<15$ M$_{\odot}$. The normalization constant, $k$, is adjusted so the mass function is continuous. Assuming a local main sequence stellar mass density of 0.034 M$_{\odot}$/pc$^3$ \citep{reid02}, the mass function in Equation 1 then yields different local spatial densities across different ranges of main sequence stellar mass. These are listed in Table 1 (column 3). In addition to spatial densities, each main sequence subpopulation has its own velocity dispersion and peculiar velocity relative to the Sun (columns 4 and 5 of Table 1), which are adopted from \citet{gar01}. Adding these velocities in quadrature and multiplying by the spatial density yields an encounter flux that can be converted into a number of encounters per Myr within 1 pc of the Sun (column 6 of Table 1). We should note outside of main sequence stars, Table 1 also includes subpopulations representing white dwarfs and giants (see last two rows). For the white dwarf density, we assume 0.004 M$_{\odot}$/pc$^{3}$ \citep{reid02}, and all other parameters of the white dwarf and giant population are taken from \citet{gar01} and \citet{rick08}. 

\begin{table*}
\centering
\begin{tabular}{c c c c c c c}
\hline
Stellar Subpopulation & Mass Range & $n$ & $\sigma_{*}$ & $v_{\odot}$  & Enc. Freq. & Total Encounters \\
 & (M$_{\odot}$) & (pc$^{-3}$) & (km/s) & (km/s) & (Myr$^{-1}$) & \\
\hline
MS & 0.05--0.34 & $8.53\times10^{-2}$ & 41.8 & 23.3 &  13.1 & 65390\\
MS & 0.34--0.58 & $1.63\times10^{-2}$ & 42.7 & 17.3 &  2.40 & 11991\\
MS & 0.58--0.74 & $6.53\times10^{-3}$ & 43.4 & 25.0 &  1.048 & 5239\\
MS & 0.74--0.86 & $3.16\times10^{-3}$ & 34.1 & 19.8 &  0.399 & 1995\\
MS & 0.86--1.02 & $2.69\times10^{-3}$ & 39.2 & 23.9 &  0.395 & 1977\\
MS & 1.02--1.2 & $1.82\times10^{-3}$ & 37.4 & 26.4 &  0.267 & 1335\\
MS & 1.2--1.5 & $1.27\times10^{-3}$ & 36.2 & 17.1 &  0.163 & 815 \\
MS & 1.5--1.9 & $5.64\times10^{-4}$ & 29.1 & 17.1 &  $6.10\times10^{-2}$ & 305\\
MS & 1.9--2.6 & $2.70\times10^{-4}$ & 23.7 & 13.7 &  $2.37\times10^{-2}$ & 118\\
MS & 2.6--5.0 & $1.08\times10^{-4}$ & 19.7 & 17.1 &  $9.02\times10^{-3}$ & 45.1\\
MS & 5.0--15.0 & $9.66\times10^{-6}$ & 14.7 & 18.6 &  $7.33\times10^{-4}$ & 3.67\\
WD & 0.9 & $4.44\times10^{-3}$ & 63.4 & 38.3 &  1.05 & 5271\\
Gi & 4 & $4.3\times10^{-4}$ & 41.0 & 21.0 &  $6.34\times10^{-2}$ & 317\\
\hline
\end{tabular}
\caption{List of the stellar subpopulation properties employed in our stellar passage algorithm. Columns from left to right are subpopulation type (MS = main sequence; WD = white dwarfs; Gi = giants), stellar mass range, number density of stars, velocity dispersion, heliocentric peculiar velocity, number of encounters per Myr with 1 pc of the Sun, and number of encounters within 1 pc of the Sun over 5 Gyrs.}
\end{table*}
Our stellar generation routine thus predicts a total stellar encounter rate of $\sim$19 passages per Myr within 1 pc of the Sun. This is near the Gaia-derived encounter rate of $\sim$22 passages per Myr \citep{bail18}. However, it is substantially higher than the 10.5 per Myr employed in \citet{rick08}. A comparison of our Table 1 and their Table 1 reveals that the difference in passage rates is almost entirely due to additional low-mass M dwarf encounters in our routine that result from our adoption of the \citet{reid02} present-day mass function. 

Table 1 also lists the total number of encounters within 1 pc over 5 Gyrs (column 7). Given these numbers, we select encounter times for each stellar subpopulation by randomly selecting times across a uniform distribution between 0 and 5 Gyrs. (Among our most massive stellar subpopulations, the encounter numbers are not whole numbers. To decide upon the inclusion of an extra encounter among these subpopulations in a given simulation, we employ a random number generator.) With subpopulation encounter times selected, we next assign masses to each passing star by sampling across the subpopulation mass range (see column 2 of Table 1) using the power-law given in Equation 1. 

Once encounter masses are set, we use the prescription given in \citet{rick08} to assign passage velocities, $V$, to individual encounters. First, a ``dispersion'' contribution, $v_*$, to the passage velocity is generated with the following equation: 

\begin{equation}
v_* = \sigma_*\sqrt{\frac{\eta_u^2 + \eta_v^2 + \eta_w^2}{3}}
\end{equation}
where $\sigma_*$ is the velocity dispersion of the stellar subpopulation, and the $\eta$ terms are three normally distributed random numbers with mean 0 and variance 1 representing orthogonal components of this velocity contribution. Next, $v_*$ is added to the Sun's peculiar velocity, $v_{\odot}$, relative to the subpopulation in the following manner to attain the stellar passage velocity, $V$:

\begin{equation}
V = \sqrt{v_{\odot}^2 + v_*^2 - 2Cv_{\odot}v_i}
\end{equation}
where $C$ is a random number between -1 and 1 to represent the cosine of a randomly oriented angle between the two passage velocity components, $v_*$ and $v_{\odot}$. Within each subpopulation, we define a maximum passage velocity, $V_{max}=3v_*+v_{\odot}$, that we should not typically expect to be exceeded. With $V$ selected, we draw yet another random number, $\xi$, uniformly between 0 and 1. If $\xi < V/V_{max}$, then the passage velocity is kept. Otherwise it is redrawn. This biases our encounter velocities to the high-end of the subpopulation's distribution, as we should expect given the relative flux contributions \citep{rick08}. 

Finally, with a velocity vector chosen, the star's initial position is selected randomly on a spherical surface of radius 1 pc surrounding the Sun. Then the velocity vector is oriented randomly, consistent with an isotropic distribution at every point on the sphere \citep{hen72}. In the interests of keeping the number of stellar encounters in each simulation set manageable, stellar passages are only included if their impact parameter is below 0.1 pc\footnote{ As we detail in subsequent sections, the effects of field star passages scale well with the passages' impulse gradients. Another approach would have been to only include passages above a certain impulse gradient. In this case, the maximum impact parameter would scale with the square root of stellar mass over stellar velocity.}. This set of simulations is designed to understand how a realistic sequence of stellar encounters could influence the planets' dynamical evolution over the Sun's remaining main sequence lifetime.

\subsection{Additional Simulations}

Our second set of 1000 simulations (named {\it control}) are designed to be a control set of simulations that do not include the effects of passing stars. These have the exact same initial conditions as {\it withstars}. In spite of the set name, we follow the same routine as \citet{kaibray24} and still do include the same stellar passage sequences from {\it withstars}, but in {\it control}, the stellar masses are reduced by a factor of 1000. 

Our next three sets of simulations are designed to provide better insights into dynamical evolution trajectories where the solar system is perturbed by an exceptionally close stellar encounter during the next 5 Gyrs. \citet{ray24} and \citet{kaibray24} both showed that the effects on the solar system scaled with a stellar encounter's impulse gradient. After analyzing the stellar encounters in our {\it withstars} simulation set, we find that the maximum impulse gradient experienced by the solar system over 5 Gyrs can vary by more than 3 orders of magnitude across our 1000 solar system realizations. Only 5\% of our systems experience a stellar passage yielding an impulse gradient greater than 0.45 m/s/au. These are the types of passages we explore in our next set of simulations: {\it strongpass}. 

To initialize {\it strongpass} simulations, we randomly choose an epoch in one of our {\it control} simulations between 0 and 5 Gyrs (since the powerful passage can occur at any point). Then stellar encounters are randomly drawn from our stellar passage routine described in the prior section until a pasage with an impulse gradient over 0.45 m/s/au is generated. Our randomly chosen {\it control} simulation is then exposed to this encounter, and the system is integrated until it reaches $t=5$ Gyrs. Since the initial epoch of each simulation is randomly selected, each integration length among these 1000 systems is unique. 

Another set of very similar simulations is also performed. This one, called {\it strongpass\_withstars}, is generated in nearly the same manner as {\it strongpass}. The difference is that instead of choosing random epochs from {\it control} simulations, we instead choose random epochs from the {\it withstars} simulations. After being exposed to the powerful passage just as in {\it strongpass}, the integrations are then continued to $t=5$ Gyrs including the rest of the previously employed stellar passages from the {\it withstars} run. This is done to see if there is any compounding effect of the weaker passages and our carefully selected single powerful passage. 

One more set of simulations, called {\it strongpass\_uniform} again studies the effects of stellar passages with impulse gradients over 0.45 m/s/au. Unlike the {\it strongpass} set, however, the impulse gradient of the strong encounter in each simulation is varied uniformly over the interval from 0.45--7 m/s/au (with 7 m/s/au being the largest impulse gradient seen in the {\it withstars} simulation set). For each {\it strongpass\_uniform} simulation, we randomly generate field star encounters using the {\it withstars} algorithm until an encounter with the desired impulse gradient is attained. The {\it strongpass\_uniform} encounters thus have a broad range of stellar masses, velocities and impact parameters even though impulse gradients vary systematically across the simulation set. This allows us to methodically study how the dynamical evolution of the solar system responds to stellar encounters of steadily (linearly) increasing impulse gradient. This final set of integrations is repeated a second time with the REBOUNDx WHFast integrator. As we will see, the stellar encounters can result in solar system instabilities, and this repeated integration is done to assess the dependence of instability rate on algorithm choice. In these repeated integrations, Pluto is not included, since we will see it is often unstable, and the most basic WHFast algorithm does not support close encounters between massive bodies. Similarly, since the basic version of WHFast does not support stellar passages, our WHFast re-integrations of the {\it strongpass\_uniform} systems begin 10 Myrs after the powerful stellar encounter (using the post-passage states of the corresponding MERCURY integrations as initial conditions). 

We should note that none of our simulations include the possibility of encounters with giant molecular clouds (GMCs), even though such encounters have certainly occurred in the past and will as well in the future. The large radius of such clouds would seemingly lead to impulse gradients that are inconsequential compared to those of stars. However, the mass distribution within clouds appears to be fractal in nature, and a denser clump of mass near the Sun could generate a dynamically significant perturbation \citep{brunfern96}. The dynamical importance of GMC encounters on the planets remains unknown at present. 

\section{Results}\label{sec:res}

\begin{figure}
\centering
\includegraphics[scale=0.42]{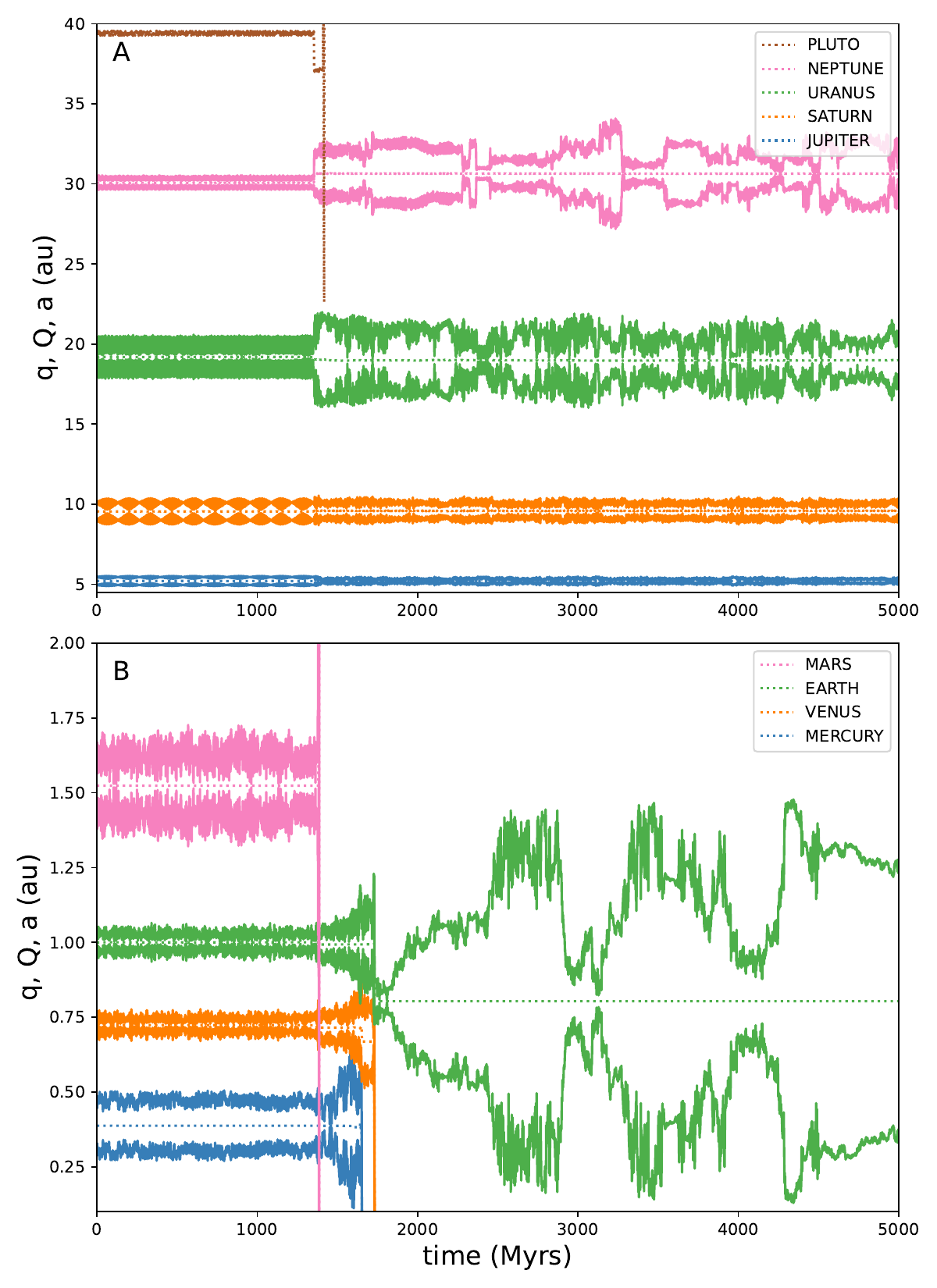}
\caption{Evolution of perihelia, aphelia, and semimajor axes vs time in a simulated solar system perturbed by passing field stars. Jupiter, Saturn, Uranus, Neptune, and Pluto are shown in Panel A. Mercury, Venus, Earth, and Mars are shown in Panel B.}
\label{fig:ExampInstab}
\end{figure}

Over the course of 5 Gyrs of evolution, five of our 1000 {\it withstars} simulations lose one or more planets. An example of one of these systems is shown in Figure \ref{fig:ExampInstab}. This simulation undergoes an exceptionally strong stellar encounter at $t=1.3$ Gyrs. At this time, a 0.25 M$_{\odot}$ star passes within 55 au of the Sun at 38 km/s. The effects of this passage on the outer solar system are obvious in Figure \ref{fig:ExampInstab}A. Here we see that the eccentricities of Uranus and Neptune are significantly excited, and both planets' semimajor axes are also perturbed. In addition, Pluto's orbit is immediately knocked out of resonance with Neptune, and it reaches the Oort cloud within 50 Myrs before ultimately being ejected $\sim$400 Myrs after the stellar passage. It is also clear that the qualitative nature of Jupiter's and Saturn's eccentricity oscillations changes immediately after the encounter.

Figure \ref{fig:ExampInstab}B shows that the outer solar system's dynamical perturbation also has significant consequences for the inner solar system. Mars is excited to a high eccentricity and ejected $\sim$30 Myrs after the stellar passage. After this point, Mercury's eccentricity increases until it collides with Venus $\sim$300 Myrs after the encounter. Finally, Venus and Earth collide with one another after another 80 Myrs of evolution. After this point, the Earth (now possessing all the mass of Venus and Mercury as well) spends large amounts of time at extreme ($>0.5$) eccentricities. Thus, this stellar encounter ultimately completely reshapes the inner solar system into a single massive, eccentric planet via one ejection and two planet-planet collisions. 

While it is already known that the solar system's internal dynamics can drive planetary instabilities in the inner solar system, there are several notable differences between instabilities of that nature compared to what is shown in Figure \ref{fig:ExampInstab}B. First, among ``internally driven'' instabilities, it seems that these events always begin with a Mercury eccentricity excitation and only rarely bleed over to Earth and Mars \citep{laskgast09}. (Typically the end result is a collision between Mercury and Venus or the Sun.) Meanwhile, in Figure \ref{fig:ExampInstab}B, it is Mars' eccentricity that is initially excited, and it is lost well before Mercury. In addition, every inner planet is either ejected or undergoes a planetary collision. Such outcomes appear to be very rare in the absence of stellar perturbations \citep{laskgast09}. Finally, because the probability of a close stellar passage is uniformly distributed in time, the types of instabilities shown in Figure \ref{fig:ExampInstab} have the potential to occur relatively quickly in the future evolution of the solar system. In contrast, internally driven instabilities typically require Gyrs of prior evolution, and the majority occur in the final Gyr of the Sun's main sequence phase \citep{abbot23}. Nearly no internally driven instabilities have been documented to occur as quickly as the ``stellar-driven'' instability depicted in Figure \ref{fig:ExampInstab} \citep{abbot23}.

While Figure \ref{fig:ExampInstab} displays unique dynamical behavior, it is not clear how common such events are relative to the more thoroughly studied internally driven instabilities. Among our 1000 {\it withstars} simulations, just five systems lose one or more planets via instability. This is more than the single system that goes unstable in our {\it control} simulations, but these are extremely small number statistics. Both sets of runs feature instability rates ({\it withstars}: $0.5^{+0.6}_{-0.3}$\%; {\it control}: $0.1^{+0.5}_{-0.08}$\%) that are lower than the expected $\sim$1\% rate expected for internally driven instabilities. Some of this could be bad statistical luck, as a larger number of systems (three in {\it control} and eight in {\it withstars}) record Mercury eccentricities above 0.65, which has been used for an instability criterion in other work \citep{jav23}. This would yield instability rates among our {\it withstars} and {\it control} systems of $0.8^{+0.8}_{-0.4}$\% and $0.3^{+0.6}_{-0.2}$\%. (The uncertainties we quote in our instability rates mark the bounds of the 95\% confidence intervals utilizing the Wilson score interval, which becomes asymmetric for very high or low probabilities.)

Another potential contributing factor to our lower rate of instability is the MERCURY hybrid integrator, which symplectically integrates systems in democratic heliocentric coordinates \citep{dun98}. We choose this integration algorithm for its ability to handle close encounters between massive bodies as well as stellar passages \citep{cham99, kaib18}. However, most prior ensembles of solar system integrations instead use Jacobi coordinates \citep{abbot23}. Starting from systems already possessing moderately high Mercury eccentricity ($e\sim0.5$), \citet{zeeb15} show that subsequent integration in Jacobi coordinates are significantly more likely to yield further increases in Mercury's eccentricity compared to the same simulations performed in democratic heliocentric coordinates. Thus, integrations performed in democratic heliocentric coordinates may underestimate the rate of Mercury instability.

Regardless of the low absolute number of instability events, there are a couple pieces of evidence supporting the idea that stars play a major role in driving the {\it withstars} instabilities and account for the higher rate of instabilities in this simulation set. First, four of our five unstable systems feature a stellar passage with an impulse gradient placing in the 96th percentile or higher of the maximum one expected over a 5 Gyr timespan. This would have a probability less than 10$^{-5}$ of occurring by chance. 

\begin{figure}
\centering
\includegraphics[scale=0.42]{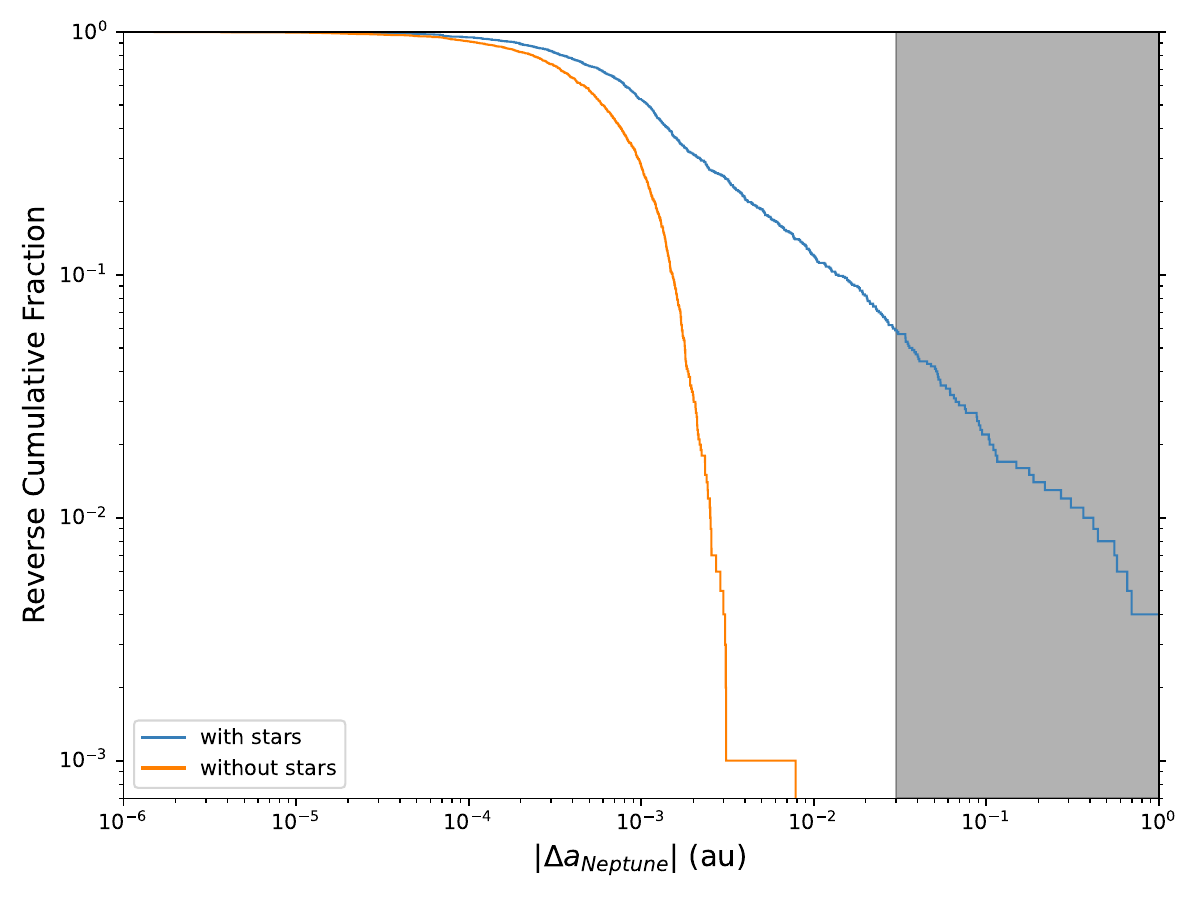}
\caption{Distribution of absolute changes in Neptune's final semimajor axis relative to its current value. Blue marks the distribution for our {\it withstars} systems, and orange marks our {\it control} systems. The shaded region corresponds to semimajor axis changes greater than 0.1\%.}
\label{fig:NepShift}
\end{figure}

Second, it has already been established in prior work that stellar passages that alter Neptune's orbital elements by more than 0.1\% enhance the solar system's instability probability by roughly one order or magnitude, or to $\sim$10\% \citep{brownrein22}. In the case of Neptune's semimajor axis, this amounts to a shift of $\pm$0.3 au. In Figure \ref{fig:NepShift}, we look at the final distribution of Neptune's semimajor axis relative to its initial value (measured in the Jacobi frame and averaged over 20 Myrs) for all 1000 of our {\it withstars} and {\it control} systems. The influence of field star passages is obvious, and the two distributions look very different. We also see that $6\pm1.5$\% of our {\it withstars} systems experience Neptunian semimajor axis shifts over $\pm$0.1\%. If we assume that these systems have a $\sim$10\% chance of instability, this means that $\sim$0.6\% of our systems (or 6 out of 1000) should experience an stellar passage-triggered instability, which is near our actual rate of 5 out of 1000. We will see in subsequent sections that the 0.6\% instability rate this simple estimate provides appears to be quite accurate. 

\subsection{Perturbations to Secular Architecture of the Giant Planets}

It is clear from Figures \ref{fig:ExampInstab} and \ref{fig:NepShift} that 5 Gyrs worth of field star passages have the potential to significantly influence the orbits of the outer planets (and, by extension, the inner planets). Prior work modeling the long-term orbital evolution of the solar system in isolation has found that while the giant planets may be chaotic, their orbital element fluctuations are nearly regular and do not rapidly diffuse due to chaos \citep{hay07, lask08}. Moreover, the giant planets' secular eigenfrequencies (and their amplitudes) are nearly fixed, analogous to dynamical metronomes \citep{hoanglask21}. 

\begin{figure}
\centering
\includegraphics[scale=0.42]{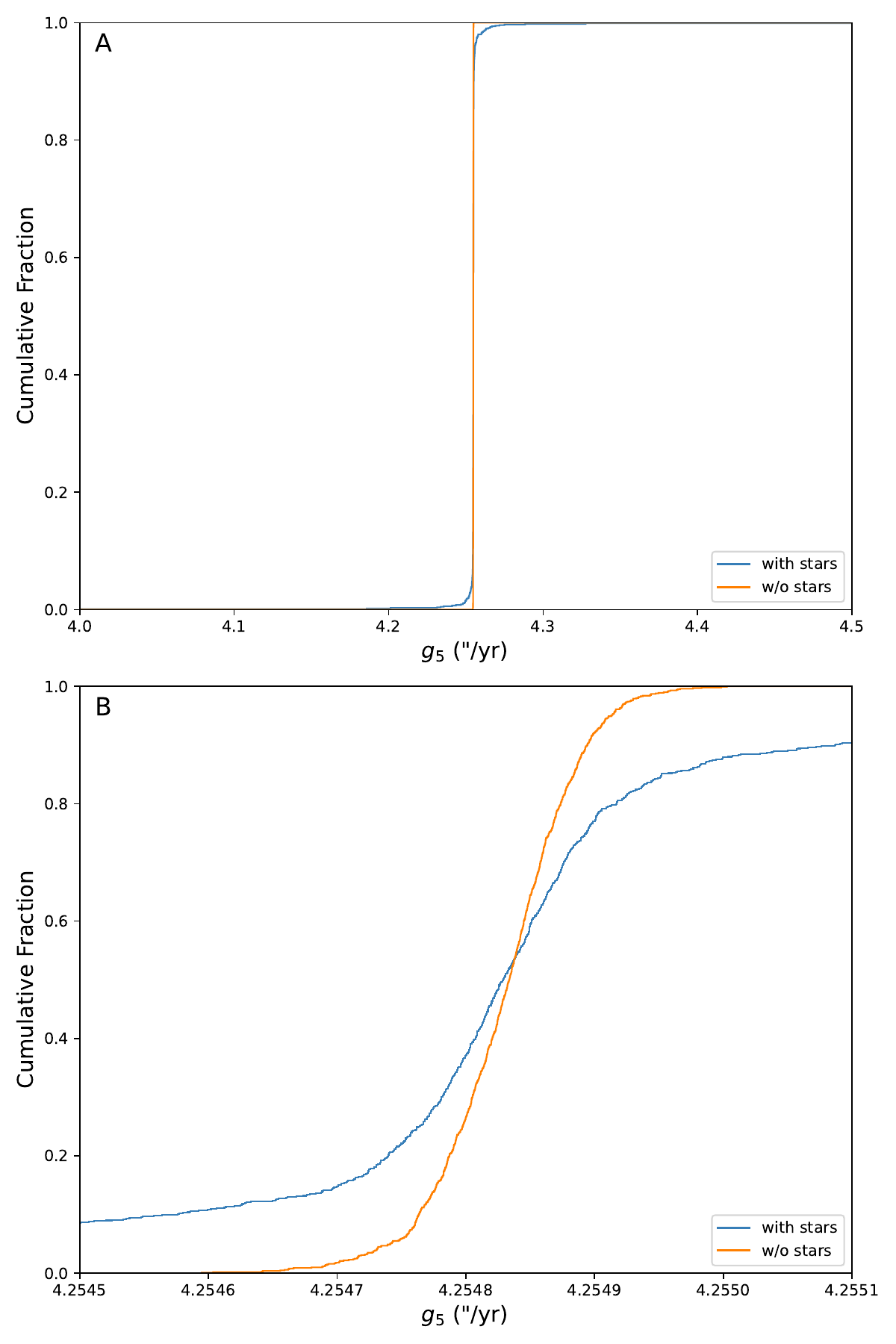}
\caption{Distribution of Jupiter's final eccentricity eigenfrequency ($g_5$) for our 1000 {\it withstars} systems (blue distribution) and our 1000 {\it control} systems (orange distribution). Panel B is a zoomed in version of Panel A.}
\label{fig:JupDiffuse}
\end{figure}

This situation may change under the influence of passing stars. In Figure \ref{fig:JupDiffuse}, we plot the distribution of eigenfrequencies associated with the precession of Jupiter's eccentricity vector in Laplace-Lagrange theory ($g_{5}$) for all 1000 of our {\it withstars} runs, and then we compare it against the same distribution measured for our {\it control} runs. (These frequencies are measured with the FMFT algorithm \citep{lask92, lask93, sidnes96, lask99}, sampling orbital elements every 1000 years for each simulation's final $\sim$8.2 Myrs.) We see in Panel A that the two distributions cluster around the same values, but the {\it withstars} distribution has wings extending $\sim$0.1 $\arcsec$/yr away from the distribution's core. These types of deviations are effectively never seen in runs performed in isolation \citep{hoanglask21}. A zoomed in view in Panel B shows that the spread of values in the {\it control} distribution is limited to a few ten-thousandths of an arc-second per year, and the standard deviation of the sample is $5.4\times10^{-5}$ $\arcsec$/yr, which is very close to the standard deviation of $\sim$$4.4\times10^{-5}$ $\arcsec$/yr predicted in \citet{hoanglask21}. This is 3--4 orders of magnitude smaller than the wings of the {\it withstars} distribution, and the standard deviation of the {\it withstars} $g_5$ values is 0.052 $\arcsec$/yr, or $\sim$1000 times that of the {\it control} sample. Thus, Jupiter's orbital element fluctuations may exhibit much less regularity on Gyr timescales than previously supposed. 

\begin{figure}
\centering
\includegraphics[scale=0.42]{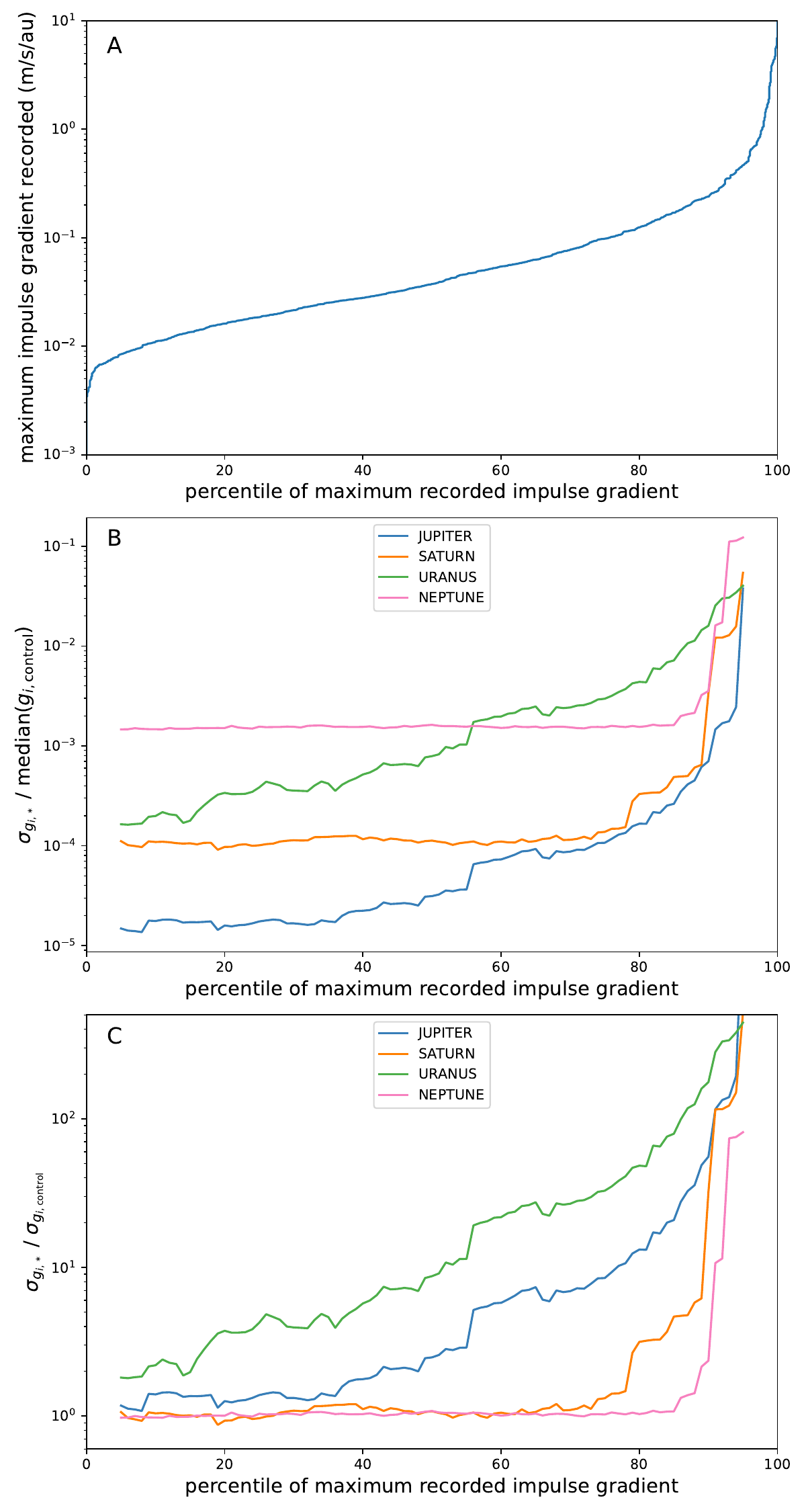}
\caption{{\bf A:} The maximum stellar passage impulse gradient of each of our {\it withstars} systems vs the impulse gradient's percentile rank among our {\it withstars} systems. {\bf B:} The standard deviation in the measured eccentricity eigenfrequencies ($g_{5-8}$) of the four giant planets as a function of our {\it withstars} systems' maximum impulse gradient percentile rank. The standard deviation is measure across a rolling window of 100 systems, or 10 percentiles of impulse gradient. Each standard deviation measurement is divided by the median value of the eigenfrequency measured in our {\it control} systems. {\bf C:} The ratio of the standard deviation of the giant planets' eccentricity eigenfrequencies in our {\it withstars} systems to the standard deviation in our {\it control} systems as a function of our {\it withstars} systems' impulse gradient percentile.}
\label{fig:gmodediffuse}
\end{figure}

Each of our 1000 {\it withstars} simulations experiences a unique set of stellar passages. However, the variations in level of external perturbation from run to run are highly concentrated among the most powerful few stellar passages. For instance, if we look at the most powerful impulse gradient that each system experiences, we find that the 95th percentile system has an impulse gradient $\sim$55 times stronger than the 5th percentile system (see Figure \ref{fig:gmodediffuse}A). When we do the same comparison with the {\it second} strongest impulse gradient, we find the difference between the 95th and 5th percentile systems drops to a factor of $\sim$16. If we do the same comparison for the fourth strongest impulse gradient, only a factor of $\sim$5.6 separates the 95th and 5th percentile systems. 

Given the huge variation in the maximum impulse gradient that each system experiences, in Figure \ref{fig:gmodediffuse}B, we plot how the standard deviation of each giant planets' final eccentricity eigenfrequency ($g_{5}$ through $g_8$) changes with the percentile of a system's maximum impulse gradient (calculating standard deviations across a rolling 10 percentile window). This figure indeed shows that the maximum impulse gradient experienced from a field star passage plays a significant role in the outer solar system's dynamical evolution. In the case of Saturn and Neptune, we see that the distribution of their eigenfrequencies remains relatively constant for systems whose maximum impulse gradient places below the $\sim$80th percentile. However, between the 80th and 99th percentiles, the standard deviation in eigenfrequency rapidly rises by about two orders of magnitude for each planet. 

In the case of Jupiter, the effects of stars begin to be seen in more weakly perturbed systems, and the eigenfrequency's standard deviation starts to rise for systems whose maximum impulse gradient is near the 50th percentile. Uranus' eigenfrequency appears to be even more sensitive to field stars. The standard deviation in eccentricity eigenfrequency grows across our entire {\it withstars} simulation set, suggesting that Uranus' secular evolution is significantly influenced under any plausible set of field star passages. Compared to the eigenfrequency diffusion in isolated solar system models, these effects can become quite large. For systems whose maximum impulse gradient is between the 80th and 90th percentile, we see a standard deviation in Uranus' eccentricity eigenfrequency of roughly 1\%. This variance is approximately 3 orders of magnitude larger than predicted in solar system models that do not consider external perturbations \citep{hoanglask21}. Although our numerical recovery of eigenfrequencies is obviously not perfectly precise, Figure \ref{fig:gmodediffuse}C shows that the standard deviations of the {\it withstars} and {\it control} simulations are nearly the same if a system's maximum impulse gradient is a lower percentile, and the two batches steadily diverge for higher percentiles. 

\begin{figure}
\centering
\includegraphics[scale=0.42]{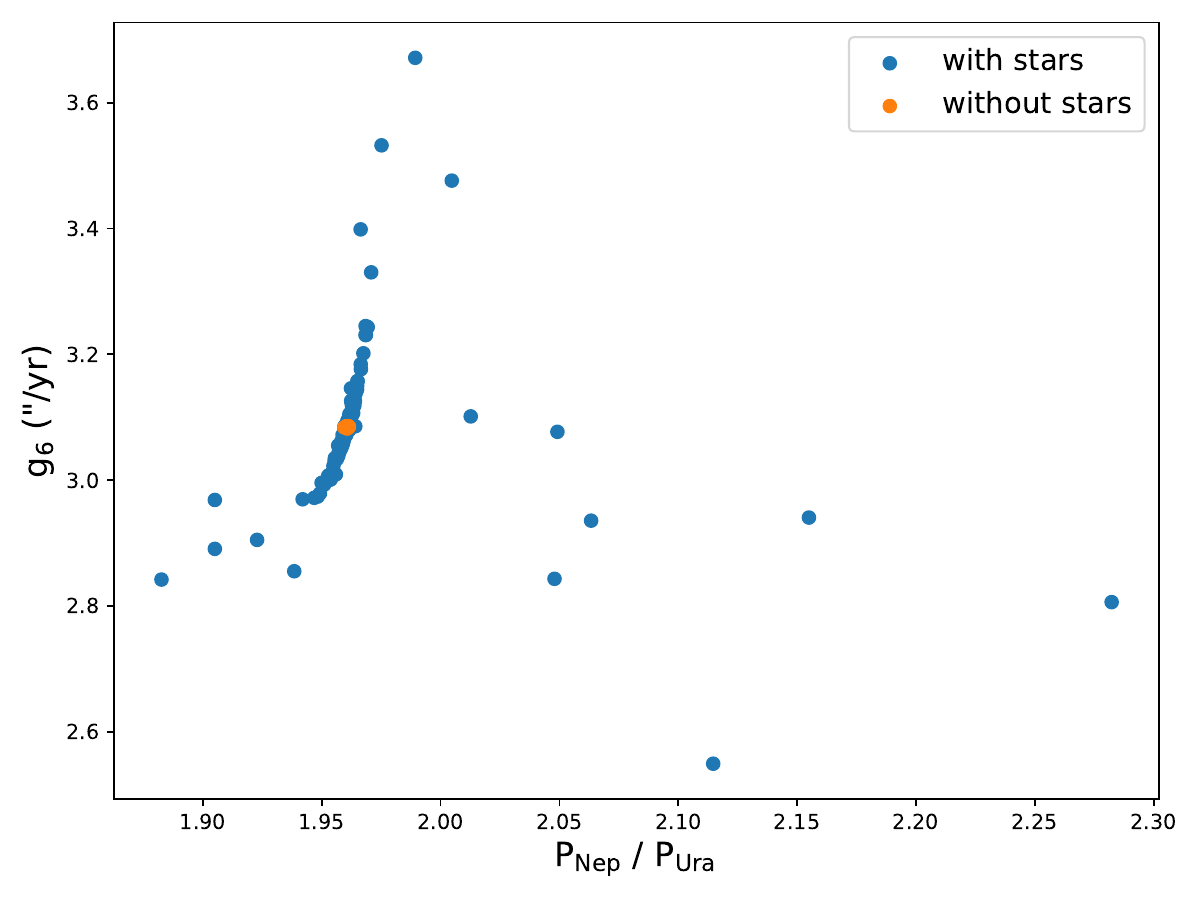}
\caption{Plot of the eigenfrequency of Uranus' eccentricity vector ($g_6$) vs the ratio of Neptune's orbital period to Uranus' orbital period. Blue points mark our 1000 {\it withstars} systems, and orange points mark our 1000 {\it control} systems. Period ratios are taken to be the median value measured in Jacobi coordinates during a system's final 20 Myrs.}
\label{fig:UResDriver}
\end{figure}

Figure \ref{fig:UResDriver} shows why it is Uranus' eccentricity eigenfrequency that is most sensitive to field star passages. In this figure we plot the measured eigenfrequency vs the ratio of Neptune's orbital period to Uranus' at the end of each of our 1000 simulations. From Figure \ref{fig:NepShift} it is already known that field stars alter Neptune's semimajor axis in many of our {\it withstars} systems, and this must in turn usually lead to a change in the ratio of Neptune's orbital period to Uranus' period. Currently, Neptune and Uranus lie very near their 2:1 mean motion resonance, or MMR, with a ratio of $\sim$1.96. The proximity to this first order resonance significantly increases the rate that Uranus precesses. Figure \ref{fig:UResDriver} shows that if stellar encounters happen to leave Uranus further from this resonance, it precesses significantly slower, but if they push Uranus even closer to the 2:1 MMR, then the precession rate accelerates. Thus, shifting the period ratio by 1--2\% can result in a nearly 10\% variation in Uranus' eigenfrequency. 

\begin{figure*}
\centering
\includegraphics[scale=0.35]{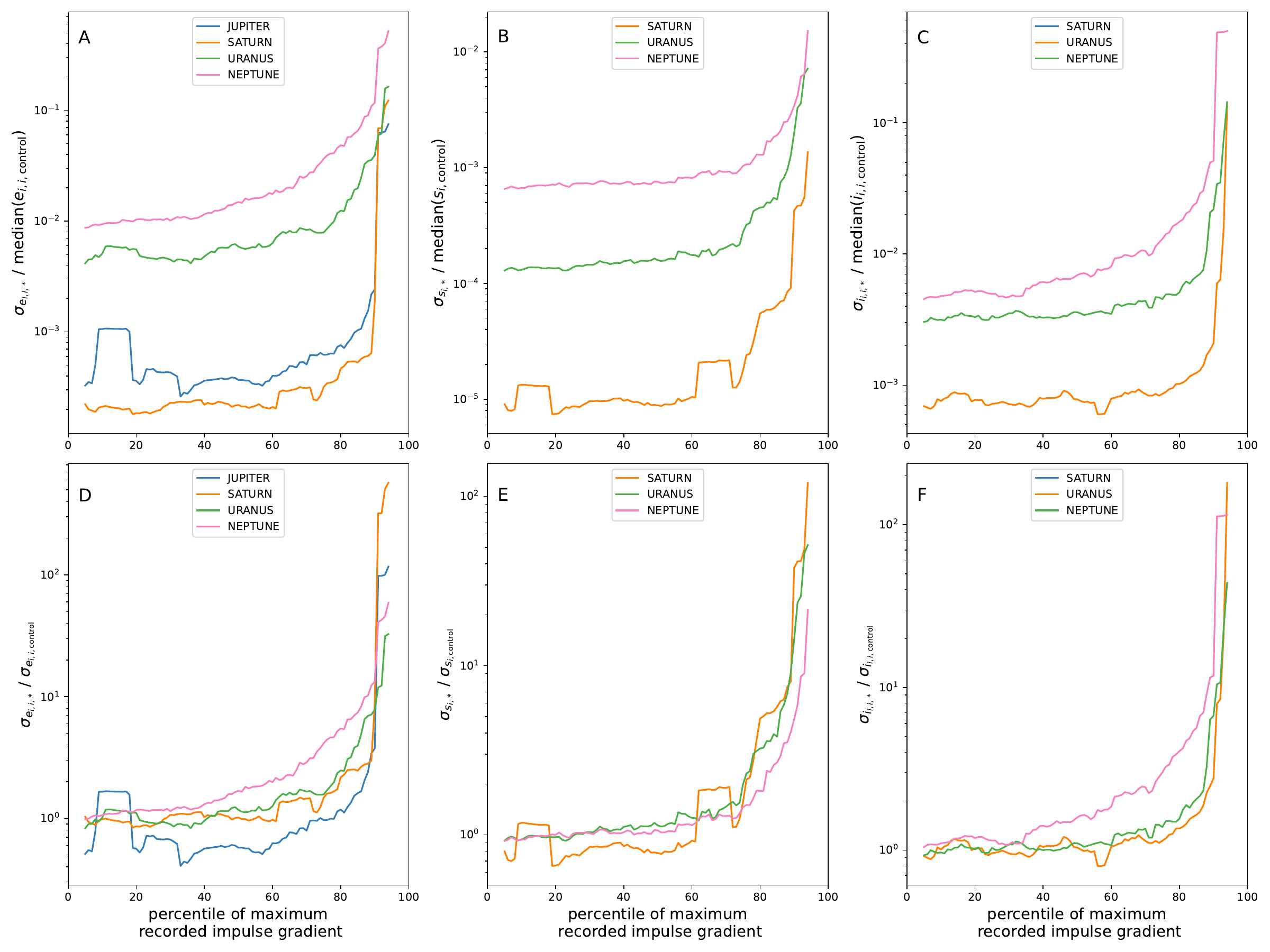}
\caption{{\bf A:} Standard deviation of the amplitude of each giant planet's eccentricity eigenfrequency ($e_{55}$ thru $e_{88}$) in our {\it withstars} systems divided by the median amplitude value in all {\it control} systems plotted against percentile of the impulse gradient of the most powerful stellar passage for our {\it withstars} systems. Standard deviations are measured across a rolling window of 100 systems (10 percentiles). {\bf B:} Same as Panel A except the standard deviation of each giant planet's inclination eigenfrequency ($s_{6}$ thru $s_{8}$) is considered. {\bf C:} Same as Panel A \& B except the standard deviation of each giant planet's inclination eigenfrequency amplitude ($i_{66}$ thru $i_{88}$) is considered. {\bf D--F:} These are similar to panel's A--C except the ratio of the standard deviation in {\it withstars} to the standard deviation in {\it control} is plotted as a function of the percentile of the systems' impulse gradient of the most powerful stellar passage.}
\label{fig:othersecdiffuse}
\end{figure*}

In the interests of completeness, Figure \ref{fig:othersecdiffuse} shows how the standard deviation of eccentricity eigenfrequency amplitudes ($e_{55}$ through $e_{88}$) varies with the percentile rank of systems' maximum impulse gradient (panels A \& D). In addition, the same is also shown for the frequency and amplitude of the giant planets' inclination/nodal regression eigenmodes as well (panels B, C, E, \& F). None of the terms appear as sensitive to field star passages as Uranus' eccentricity eigenfrequency ($g_7$), yet the standard deviation of all amplitudes and frequencies eventually climb rapidly as we reach systems whose maximum impulse gradients reach the 80th--90th percentile. Of particular note is the amplitude of Neptune's eccentricity eigenmode, $e_{88}$. This is effectively the eccentricity Neptune would have in the absence of forcing from other planets. For a system whose maximum impulse gradient is near the median, we find a standard deviation of 1--2\%. By the time we reach $\sim$85th percentile systems, this standard deviation has grown to nearly 10\%, suggesting that passing field stars can and could have been a significant contributor to Neptune's eccentricity \citep{brun93}. 

\subsection{Stability of Pluto}

\begin{figure}
\centering
\includegraphics[scale=0.42]{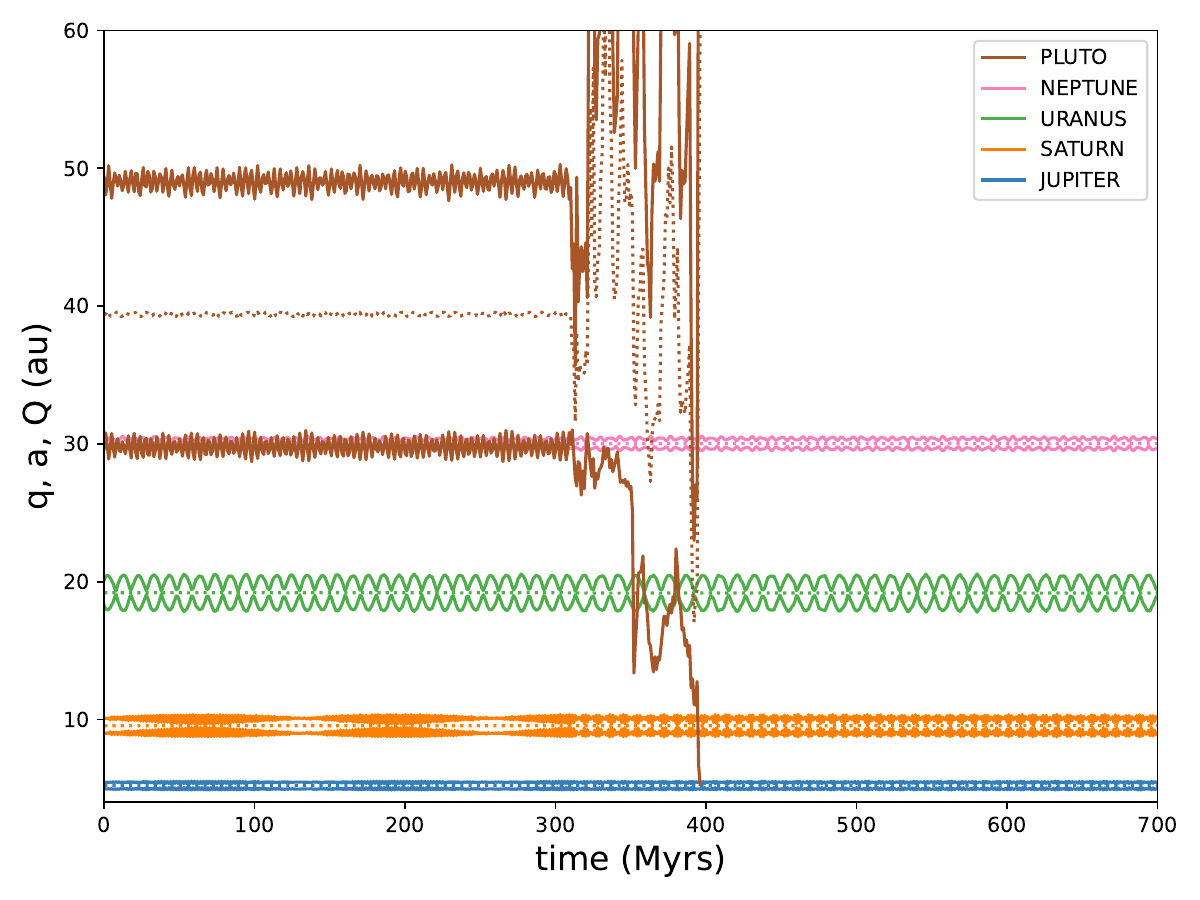}
\caption{Plot of the perihelion, aphelion, and semimajor axis of Pluto, Neptune, Uranus, Saturn, and Jupiter for one {\it withstars} system as a function of time. Pluto is ejected from the system after 397 Myrs of evolution.}
\label{fig:plutoinstab}
\end{figure}

If the giant planets' orbital evolution can be altered through close field star passages, the same must be true for Pluto as well. We have already seen in Figure \ref{fig:ExampInstab} at least one case where Pluto was destabilized by a stellar passage. However, this is far from the only example. Another example is shown in Figure \ref{fig:plutoinstab}. In this case, a 0.3 M$_{\odot}$ star moving at 66 km/s passes within 138 au of the Sun at 310 Myrs into the simulation. Unlike the prior instability from Figure \ref{fig:ExampInstab}, this encounter is not strong enough to visibly alter the orbital evolution of the outer planets, and this system retains all eight planets for the entire 5-Gyr integration. However, Pluto is easily dislodged from its 3:2 MMR with Neptune. After scattering off of the giant planets for almost 100 Myrs, it is finally ejected. 

In total, 39 out of our 1000 {\it withstars} simulations lose Pluto (nearly always through ejection) over the course of their 5-Gyr integrations. Prior works studying Pluto's orbital evolution have concluded it to be 100\% stable on these timescales \citep{kinnak96, itotan02, malito22}, but our results imply that it actually has a $3.9^{+1.3}_{-1.0}$\% chance of instability over the Sun's remaining main sequence lifetime. This is purely due to the possibility of close field star passages, and it is substantially higher than the $\sim$1\% instability probability estimated for Mercury among isolated solar system models. 

\begin{figure}
\centering
\includegraphics[scale=0.42]{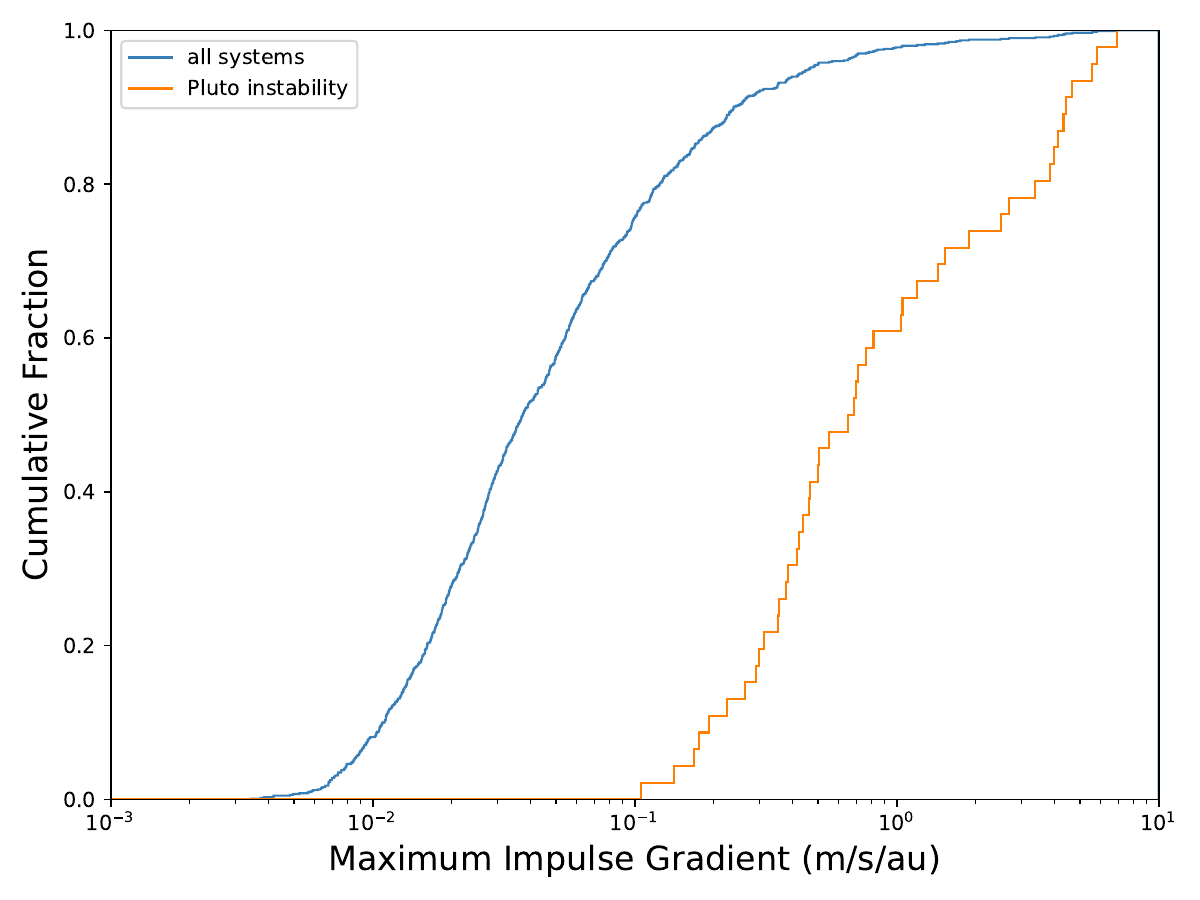}
\caption{Distribution of the maximum impulse gradients due to stellar passages that each system experiences over 5 Gyrs of evolution. Blue marks the distribution of all 1000 {\it withstars} systems. Orange marks the distribution for the 39 {\it withstars} systems that lose Pluto.}
\label{fig:plutoloss}
\end{figure}

Figure \ref{fig:plutoloss} shows the distribution of maximum impulse gradients experienced by {\it withstars} systems that lose Pluto. This is compared to the overall distribution of {\it withstars} systems' maximum impulse gradients, and we see that the distributions are quite different. The median value for maximum impulse gradient of a Pluto-losing system is $\sim$0.7 m/s/au. This is $\sim$20 times higher than the median value for all of our {\it withstars} systems. These Pluto-destabilizing passages have a median impact parameter of 115 au (with the most distant beyond 400 au), a median velocity of 47 km/s, and a median stellar mass of $\sim$0.3 M$_{\odot}$. 

\begin{figure}
\centering
\includegraphics[scale=0.42]{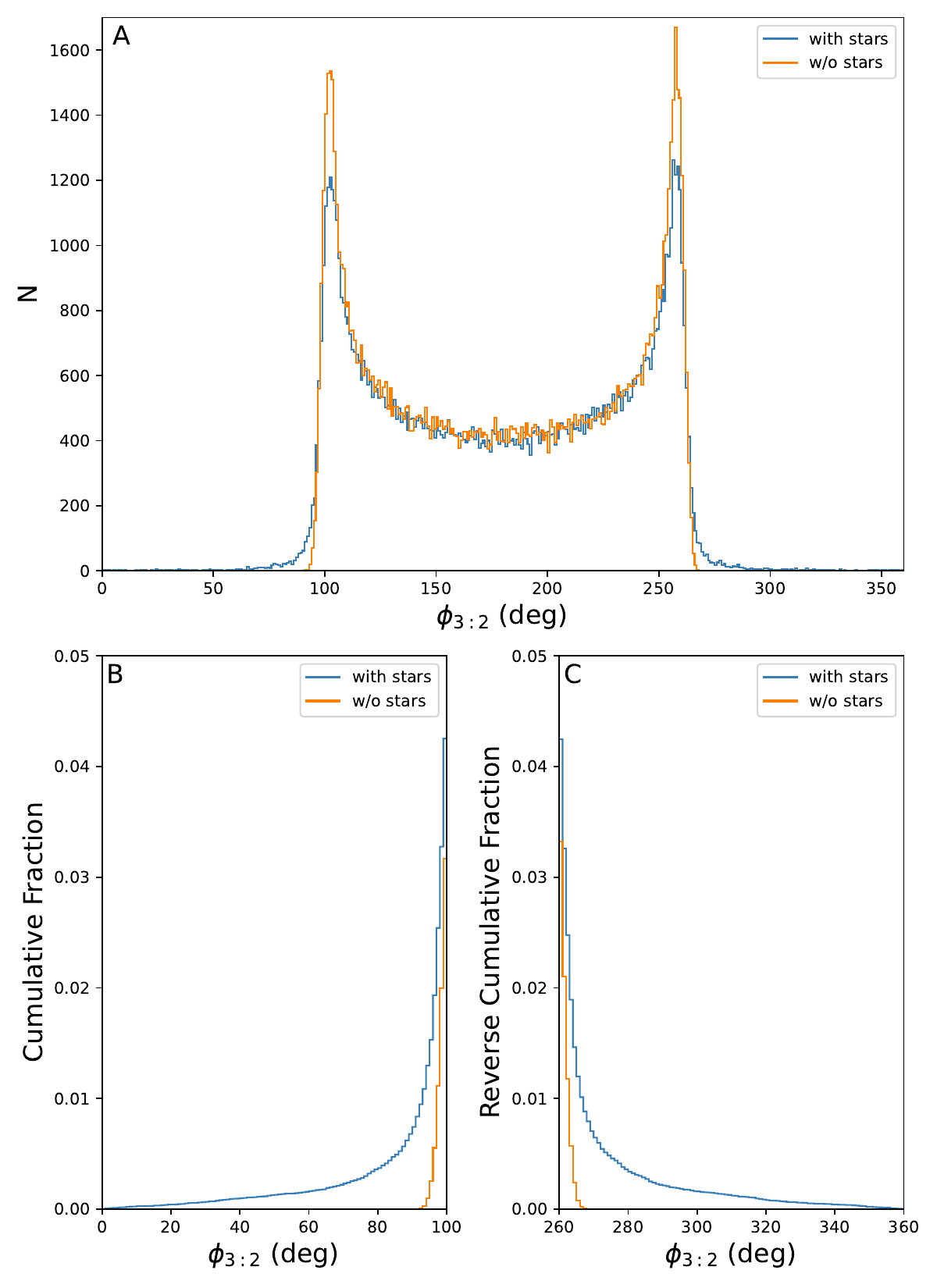}
\caption{{\bf A:} Final distribution of the 3:2 MMR resonant angle between Pluto and Neptune after 5 Gyrs of evolution. Blue marks the distribution among our {\it withstars} systems, and orange marks the distribution among our {\it control} systems. {\bf B:} A zoom-in on the left side of panel A. {\bf C:} A zoom-in on the right side of panel A. }
\label{fig:plutodiff}
\end{figure}

Only considering systems that have completely lost Pluto may underestimate the probability of major orbital changes for Pluto, however. There may also be cases in which passing stars compromise its resonant configuration with Neptune without actually ejecting it. To gauge the prevalence of this, we sample Pluto's 3:2 resonant angle with Neptune every Myr for the last 20 Myrs of each simulation. The distribution of resonant angles is shown in Figure \ref{fig:plutodiff} for both our {\it withstars} and {\it control} simulations. One can see that the resonant angle libration is neatly confined between $\sim$95$^{\circ}$ and $\sim$265$^{\circ}$ for our {\it control} simulations. This confinement of resonant angle is critical for Pluto to avoid close encounters with Neptune \citep[e.g.][]{malito22}. The {\it withstars} systems still follow the same general distribution of resonant angles as the {\it control} systems. However, there is clearly a population at the level of 1--2\% (see Panels B \& C) whose resonant angle values are never seen in our {\it control} systems. Taken on top of the $\sim$4\% of systems that actually lose Pluto, we conclude that Pluto's current resonant configuration with Neptune has perhaps a 94--95\% chance of remaining stable for the next 5 Gyrs. 

Pluto is notable as the largest object in the Kuiper belt, and many prior works have been devoted to studying its dynamical evolution \citep[e.g.][]{kinnak96, itotan02, malito22}. Our results here show that its stability is not guaranteed, and this finding may extend to other Kuiper belt objects as well. For instance, Orcus is another dwarf planet that occupies the 3:2 resonance with Neptune \citep{brown04b}. Its eccentricity and inclination are comparable to Pluto, and it may also be susceptible to destabilization via close stellar passages. The stability of known Kuiper belt objects in the presence of exceptionally strong stellar encounters should be undertaken to better understand the possible dynamical evolution of the Kuiper belt.

\subsection{Stability of the Planets}

\begin{figure}
\centering
\includegraphics[scale=0.42]{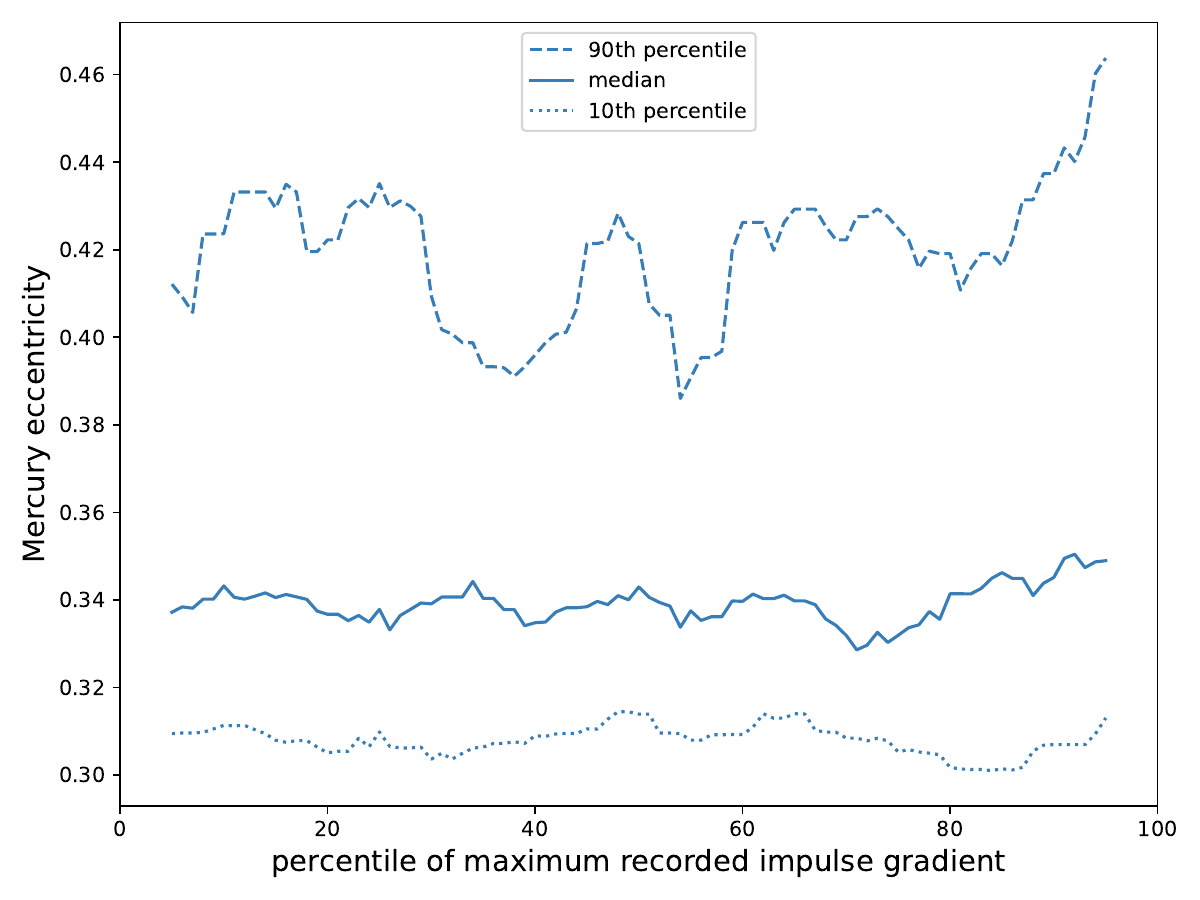}
\caption{The 90th, 10th, and 50th percentile of the maximum eccentricity recorded for Mercury among our {\it withstars} systems is plotted against the percentile of the systems' maximum impulse gradient experienced from stellar passages. The percentiles of maximum eccentricity are measured across a rolling window of 100 systems (10 percentiles).}
\label{fig:MercEcc}
\end{figure}

As evidenced in Figure \ref{fig:ExampInstab} and noted in prior work \citep{brownrein22}, stellar passages can clearly alter the stability of the Sun's planets. However, their overall dynamical significance relative to the planets' internal dynamical processes (which can also lead to instability) is unknown still. To begin understanding this, we first turn to the evolution of Mercury, since most internally driven instabilities begin with an excursion of Mercury's eccentricity and end with a collision between it and Venus or the Sun \citep[e.g.][]{laskgast09}. In Figure \ref{fig:MercEcc}, we measure the maximum eccentricity that is recorded for Mercury over each of our 1000 {\it withstars} 5-Gyr integrations. Then we plot the 10th, 50th, and 90th percentiles of the systems' maximum Mercury eccentricity value as a function of the percentile of the systems' maximum stellar impulse gradient experienced during their integrations. (Mercury maximum eccentricity percentiles are evaluated across a rolling subgroup of 100 systems ordered by their maximum stellar impulse gradient.) 

In Figure \ref{fig:MercEcc}, we see that the 10th percentile and the median of systems' maximum Mercury eccentricity do not display an obvious by-eye trend with increasing stellar impulse gradient. However, there is a notable feature when we look at the 90th percentile value of systems' maximum Mercury eccentricity. Here we see that for stellar impulse gradients below the 85th percentile ($<0.17$ m/s/au), the 90th percentile of Mercury's maximum eccentricity fluctuates around 0.42 by $\pm$0.02 with no clear trend. However, above 85th percentile stellar impulse gradients, the 90th percentile of Mercury's maximum eccentricity is higher than any other subgroup of systems with weaker stellar encounters, and it continually increases among systems that experience still more powerful encounters. The magnitude of this trend toward higher maximum Mercury eccentricity is not much larger than the random fluctuations occurring among systems with weaker encounters, but it does appear to be statistically significant. To show this, we separate our {\it withstars} systems into two groups: those whose maximum stellar impulse gradient ranks 85th percentile or higher and those that do not. We then use a K-S test to compare the distribution of maximum Mercury eccentricities recorded among the first group of systems (strongly perturbed) with those of the second group (weakly perturbed). This test returns a $p$-value of 0.003, indicating we can reject the null hypothesis with 3-$\sigma$ confidence. 

Moreover, among our {\it withstars} systems that exhibit the most extreme Mercury orbital evolution, it appears that stellar perturbations are often an important driver. As mentioned before, 5 of our {\it withstars} systems become unstable, and 4 of them feature stellar passages whose maximum impulse gradients rank 96th percentile or higher among all of our {\it withstars} systems. Instead of looking at systems that pass through a full instability, we can also look at systems where Mercury's eccentricity exceeds 0.65 at some point, as this has also been employed as a proxy for eventual instability \citep{jav23}.  In this case, we have eight systems attaining this eccentricity, and five of them again feature encounters with impulse gradients ranking 96th percentile or higher. The probability of having five (or more) such powerful encounters in eight random selections is only 5 $\times10^{-6}$ according to binomial statistics. 

To better understand the probability and nature of stellar-driven instabilities, we next turn to our {\it strongpass} set of simulations. All 1000 of these runs feature a stellar encounter that has an impulse gradient so large ($>0.45$ m/s/au) that such an encounter only has a 5\% probability of occurring in the next 5 Gyrs. These encounters are applied at random times to our {\it control} systems, and they are therefore the only significant external perturbation delivered to each system. Among our 1000 {\it strongpass} systems, 125 lose one or more planets before $t=5$ Gyrs. If we fold this $12.5\pm2.2$\% instability rate into the 5\% probability of such a powerful stellar passage occurring, we get a probability of $0.63\pm0.11$\% that the solar system's planets will be destabilized by a stellar encounter. 

In addition to our {\it strongpass} simulations, we can also look at our {\it strongpass\_withstars} simulations. These 1000 systems are generated in the same fashion as {\it strongpass}, but they utilize random epochs of our {\it withstars} systems instead of the {\it control} systems. Each system therefore undergoes many other field star passages besides the exceptionally strong one prescribed in {\it strongpass}, and in this way, they can explore the effects of cumulative weaker passages. The cumulative effects of weaker passages do not appear to decrease solar system stability, as this set of simulations actually features a somewhat {\it lower} instability rate at $10.0\pm2.0$\%.

Both simulation sets of simulations feature an instability rate that is roughly 1$\sigma$ away from the averaged instability rate of $11.3\pm1.5$\%. Folding in the fact that every system is exposed to a stellar passage that only has a 5\% chance over the next 5 Gyrs, we estimate that there is a $0.56\pm0.08$\% chance that the solar system's planets will be destabilized by a future field star encounter. This is only marginally lower than the probability for an internally driven instability amongst the solar system's giant planets, which is roughly 0.8--1\% \citep{abbot23}. Thus, destabilization via stellar passages is a major (and underexplored) dynamical pathway in our solar system's future evolution, and it increases our planets' probability for instability by a factor of $\sim$50--80\%.

\begin{figure}
\centering
\includegraphics[scale=0.42]{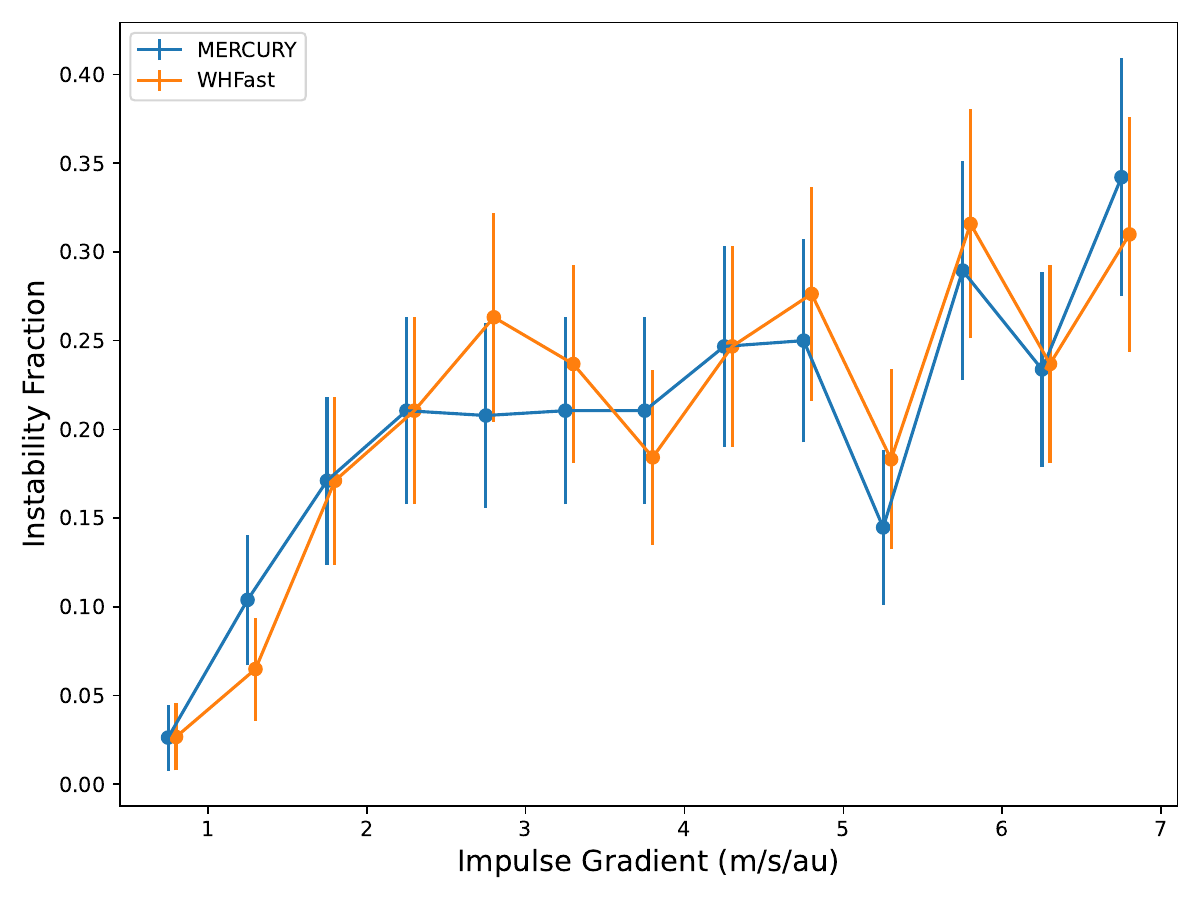}
\caption{The fraction of unstable systems (loss of one or more planets) is plotted against the stellar passage impulse gradient to which systems are exposed. Blue marks our {\it strongpass\_uniform} systems integrated with the MERCURY code, and orange marks the same systems integrated with WHFast. Error bars mark 1$\sigma$ Poisson uncertainties.}
\label{fig:ImpGradInstab}
\end{figure}

To better understand how the stellar-driven instability probability changes with stellar passages of increasing impulse gradient, we consult our {\it strongpass\_uniform} simulations. These simulations are initialized in the same way as {\it strongpass} except that the 1000 stellar passages' impulse gradients are uniformly spaced between 0.45 m/s/au and 7 m/s/au (the highest impulse gradient seen in {\it withstars}). In Figure \ref{fig:ImpGradInstab}, we plot the fraction of systems that lose one or more planets as a function of stellar impulse gradient. As can be seen, among our weakest encounters, the instability rate is just under 5\%. However, as impulse gradient increases from $\sim$1 m/s/au to 6--7 m/s/au, the instability probability increases to $\sim$30\%. Because solar system instability rates display some variance with integration algorithm \citep{zeeb15}, we confirm these instability probabilities by reintegrating the post-passage portions of our simulations with the REBOUNDx WHFast integrator. The results of these reintegrations are also displayed in Figure \ref{fig:ImpGradInstab}, and we see that the instability rates of the two simulation sets are largely similar, giving us further confidence in our simulation results.

\begin{figure}
\centering
\includegraphics[scale=0.42]{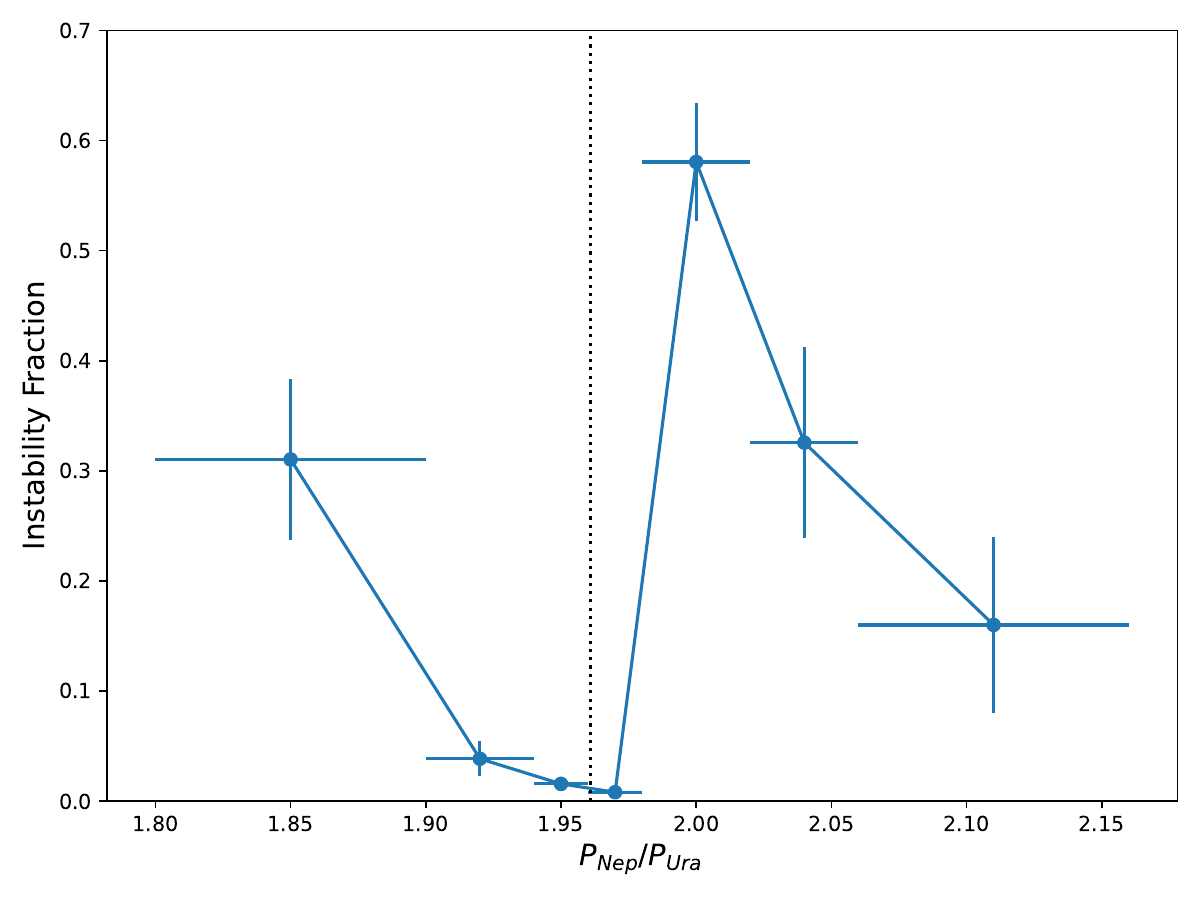}
\caption{The fraction of unstable {\it strongpass} and {\it strongpass\_withstars} systems (loss of one or more planets) is plotted against the ratio of Neptune's orbital period to Uranus' orbital period. Orbital periods are taken to be the median value during the first 100 Myrs after each system is exposed to a powerful encounter. The dotted vertical line marks the period ratio in the modern solar system. The vertical error bars mark 1$\sigma$ Poisson uncertainties, and the horizontal error bars mark the period ratio range over which systems are drawn for each data point.}
\label{fig:UNRatInstab}
\end{figure}

It has already been noted in prior work that if a stellar encounter induces a Neptunian semimajor axis change above 0.1\% then the instability probability of the Sun's planets is increased by an order of magnitude \citep{brownrein22}. Because of this, we also study how the instability rate changes as a function of the outer planets' architecture, and we find that Uranus' and Neptune's proximity to their 2:1 MMR plays a prominent role in planetary instabilities driven by stellar passages. In Figure \ref{fig:UNRatInstab}, we plot the instability probability of our {\it strongpass} and {\it strongpass\_withstars} systems as a function of the ratio of Neptune's period to Uranus' period measured immediately after each system's powerful stellar passage. (Since the additional, weaker field star passages of {\it strongpass\_withstars} simulations do not enhance instability rate, we co-add the results of {\it strongpass} and {\it strongpass\_withstars} for the remainder of this paper.) In the modern solar system, this period ratio is currently 1.96. We see that if the ratio is perturbed to 1.95 or 1.97, the instability rate is quite low ($\sim$1\%). When the period ratio is perturbed down to 1.92 (a shift of -0.04), the instability rate still remains modest at $\sim$3.8\%. However, when an equal size shift happens in the opposite direction to a period ratio of 2.00, the instability rate skyrockets to 58\%! Even stronger perturbations to the Neptune-Uranus period ratio result in still-high instability rates of $\sim$30\%, but they are lower than the systems perturbed to near the location of the 2:1 MMR. In total, only $\sim$10\% of our systems are perturbed to a Neptune-Uranus period ratio between 1.98 and 2.02. However, these systems actually generate over half (53\%) of all the instabilities seen among our {\it strongpass} and {\it strongpass\_withstars} systems. Thus, the proximity of Neptune and Uranus to their 2:1 MMR appears to significantly compromise the solar system's stability. 

\begin{figure}
\centering
\includegraphics[scale=0.42]{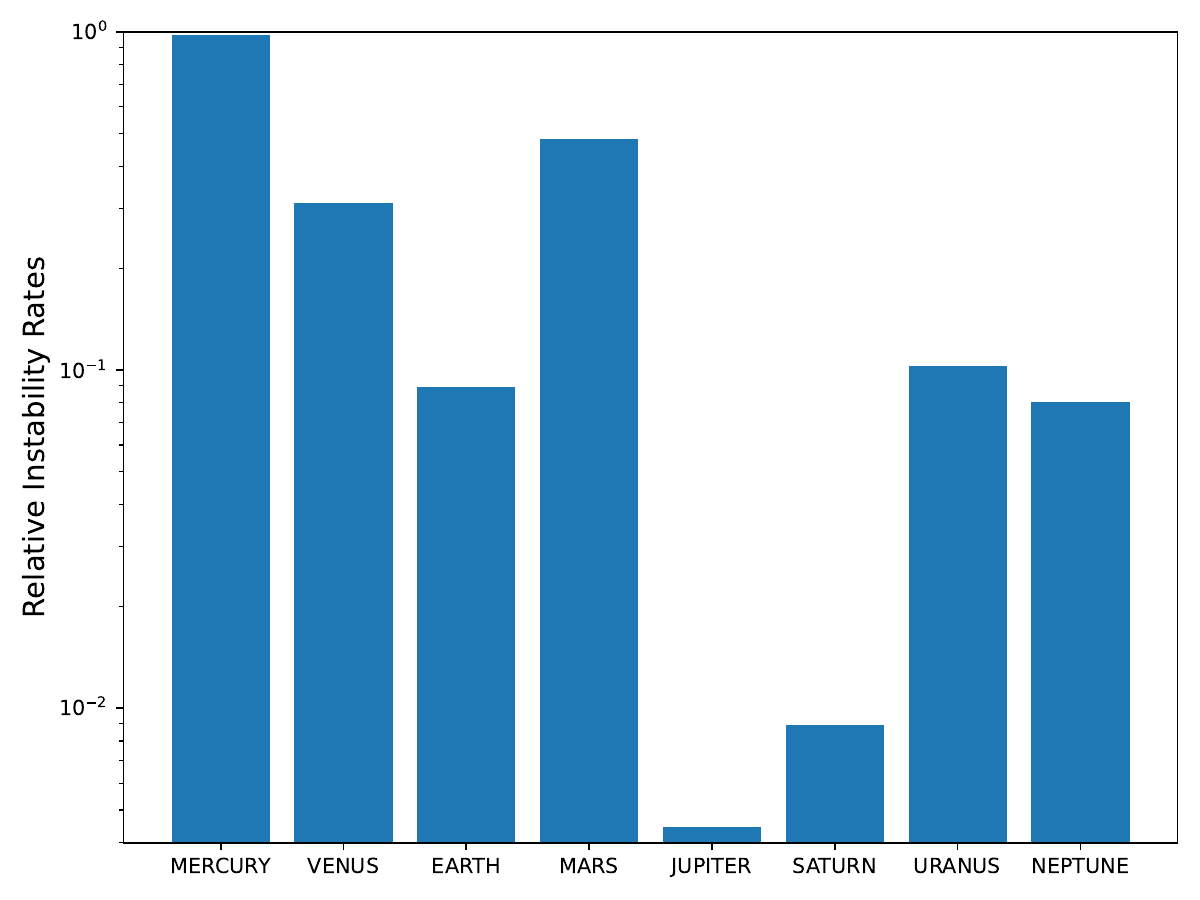}
\caption{The relative rate that planets are lost among our {\it strongpass} and {\it strongpass\_withstars} systems that lose at least one planet. A rate of 1 for a planet would indicate that all unstable systems lose that particular planet.}
\label{fig:RelativeInstabRates}
\end{figure}

Using the 225 unstable systems in the {\it strongpass} and {\it strongpass\_withstars} simulation sets, we can also determine all eight planets' relative instability rates. These are shown in Figure \ref{fig:RelativeInstabRates}. We see that Mercury is overwhelmingly likely to be lost in the event of a stellar-driven instability (most often through collision with Venus or the Sun). There are only four systems that lose a planet without losing Mercury (two feature ice giant losses and the others feature Mars losses). In the other $98.2^{+1.1}_{-2.7}$\% of unstable systems, Mercury is lost if any planets are lost. The next least stable planet is Mars. In $48\pm6.5$\% of unstable systems, Mars is eventually lost (through roughly equal probability channels of ejection, solar collision, or collision with Venus or Earth). This contrasts strongly with internally driven instabilities, which typically begin with Mercury's loss and are limited to just Mercury's loss \citep{laskgast09}. Stellar-driven instability also often leads to the loss of Venus as well, with $32\pm6$\% of unstable systems losing Venus (typically via collision with Earth). Thus, rather than just removing Mercury, stellar-driven instabilities can often completely reshape the inner solar system's architecture. The rest of the planets are more robust to stellar-driven instabilities. Earth, Uranus, and Neptune all have a $\sim$$10\pm4$\% chance of being lost in the event of an instability. (We should note, however, that Earth often suffers planetary collisions in instabilities during the loss of other inner planets.) Only two of our 225 unstable {\it strongpass} and {\it strongpass\_withstars} systems lose Saturn (both via ejection), and one ejects Jupiter. 

\begin{figure}
\centering
\includegraphics[scale=0.42]{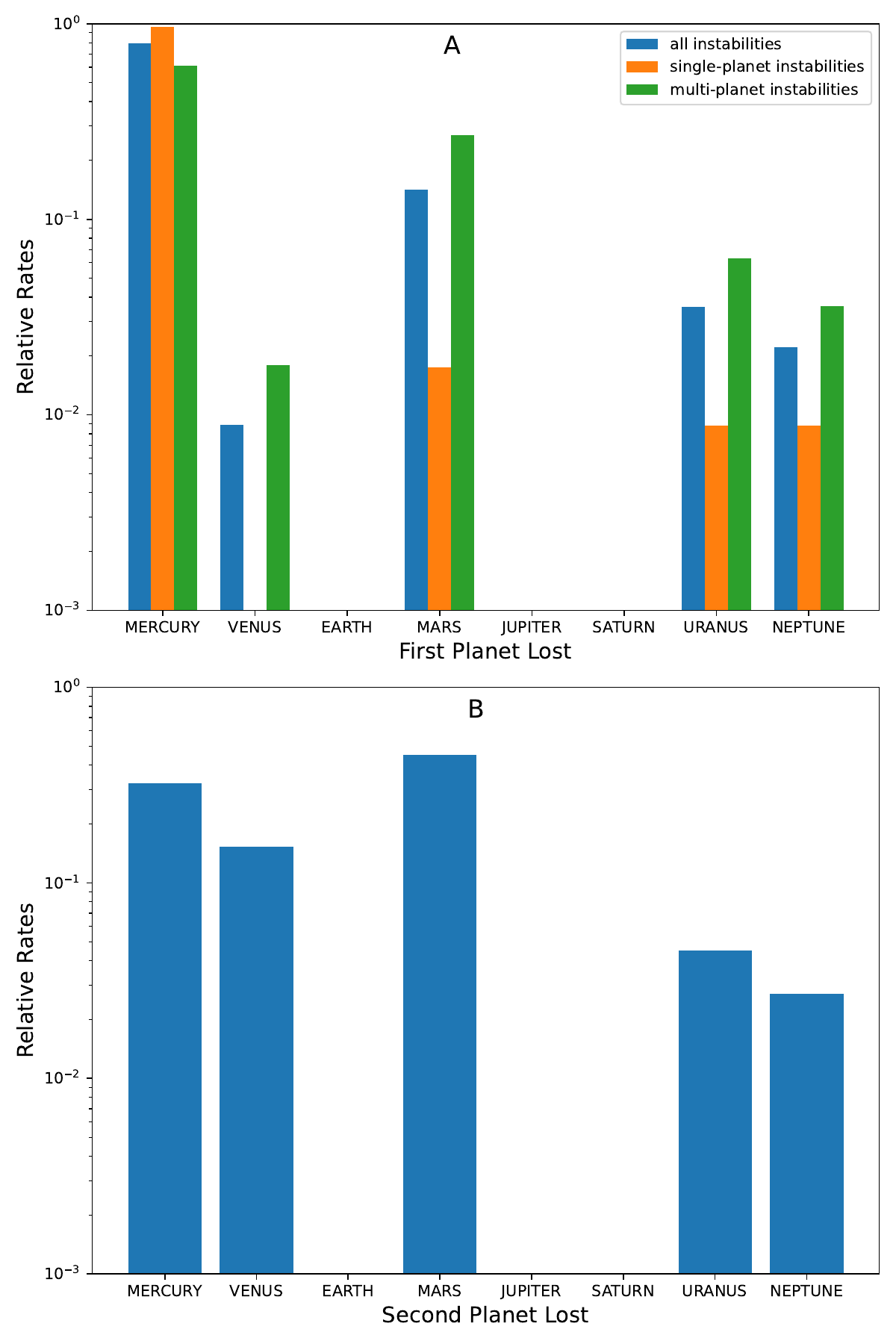}
\caption{{\bf A:} Among our 225 unstable {\it withstars} and {\it strongpass\_withstars} systems, the relative rate of which planet is the first lost is plotted. This is done for all unstable systems ({\it blue}), systems that lose just one planet ({\it orange}), and systems that lose more than one planet ({\it green}). A rate of 1 for a planet would indicate that all unstable systems lose that particular planet. {\bf B:} Among our 111 systems that lose more than one planet, the relative rate of which planet is the second one lost is plotted.}
\label{fig:PlanetLossStats}
\end{figure}

In addition, we can use our 225 unstable {\it strongpass} and {\it strongpass\_withstars} systems to study where stellar-driven instabilities begin. In Figure \ref{fig:PlanetLossStats}A, we plot which planet is the first one lost from our unstable systems. We see that, like internally driven instabilities, it is most common to lose Mercury before any other planet. This occurs in $79\pm5$\% of unstable systems. However, a significant fraction ($14^{+5}_{-4}$\%) of unstable systems, instead begin with the loss of Mars. Another $5.8^{+3.8}_{-2.4}$\% of instabilities initiate with the loss of an ice giant, and one instance begins with a Venus loss. 

Unlike internally driven instabilities, losses of multiple planets occur often among stellar-driven instabilities. Stellar-driven instabilities are nearly evenly split between those that lose just one planet ($50.2\pm6.5$\%) and those that lead to the loss of more than one planet ($49.8\pm6.5$\%). If we restrict ourselves to single-planet instabilities, we find they nearly all ($97^{+2.2}_{-5.2}$\%) involve Mercury. The only exceptions are two systems featuring a Mars loss and two featuring an ice giant loss. The majority of multi-planet instabilities also still begin with Mercury's loss, but the fraction is lower at $61\pm9$\%. Most of the remaining multi-planet instabilities ($27^{+9}_{-8}$\%) begin with the loss of Mars. Among multi-planet instabilities, Figure \ref{fig:PlanetLossStats}B shows the second planet that is lost among systems that lose more than one planet. Among instabilities that begin with Mercury, most ($71^{+9}_{-12}$\%) then proceed to the loss of Mars. Nearly all ($96^{+2}_{-5}$\%) multi-planet instabilities lead to the loss of Mars. Given the nearly even split between single- and multi-planet instabilities, this means that Mercury is only $\sim$twice as unstable as Mars in the context of stellar-driven instabilities. 

In addition to the prevalence of multi-planet losses, stellar-driven and internally driven instabilities should differ in their timing as well. In the case of internally driven instabilities, several Gyrs of prior orbital evolution is normally necessary to yield an instability, and the probability of internally driven instability after 5 Gyrs of evolution is nearly an order of magnitude larger than the internally driven instability probability after 3 Gyrs of evolution \citep{abbot23}. Meanwhile, the probability of an instability-inducing stellar encounter occurring is effectively constant (but small) with time. If stellar-driven instabilities develop over timescales that are short relative to the solar system age, then their probability should be nearly uniformly distributed over the next 5 Gyrs.

\begin{figure}
\centering
\includegraphics[scale=0.42]{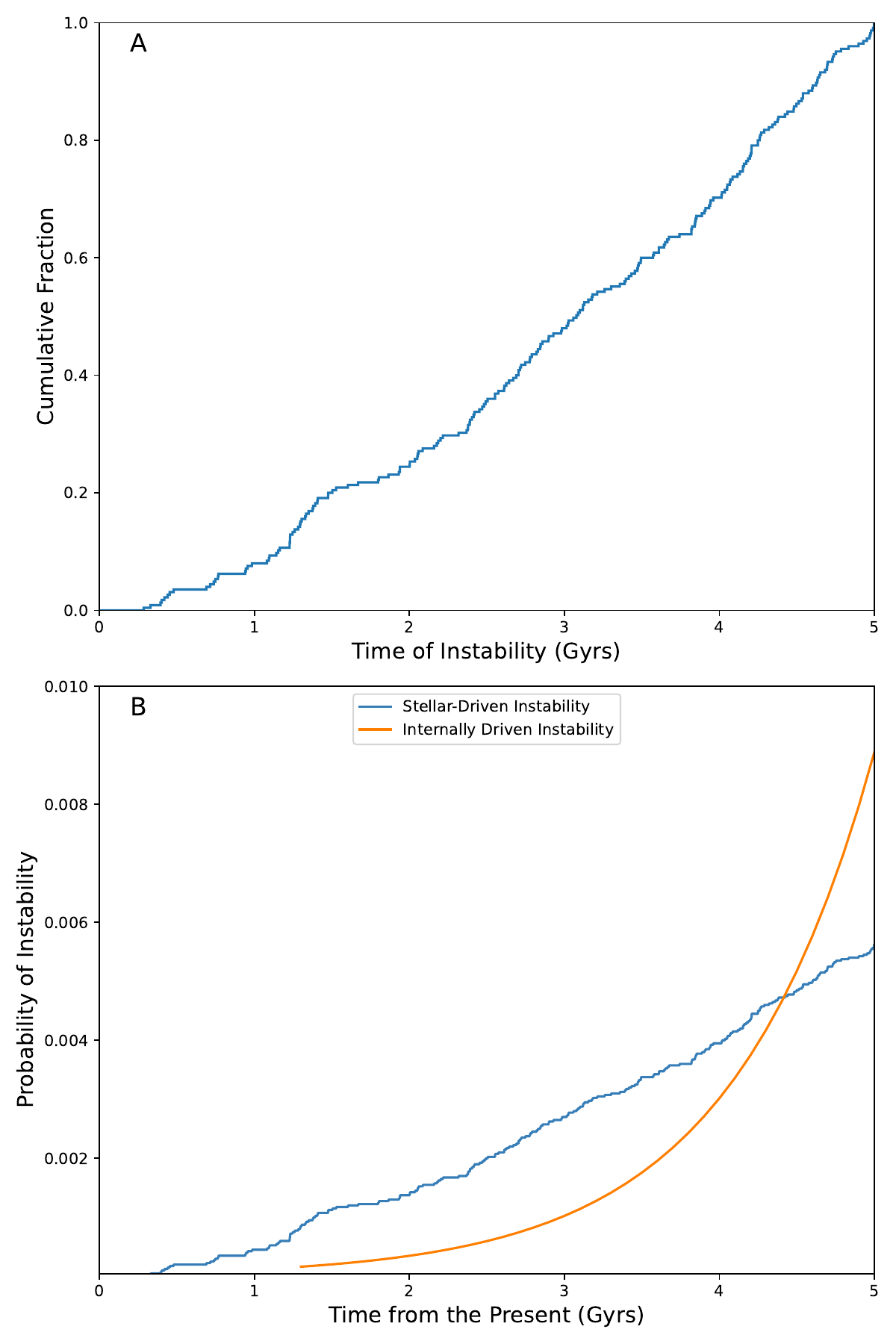}
\caption{{\bf A:} The cumulative fraction of {\it strongpass} and {\it strongpass\_withstars} systems that become unstable with time. {\bf B:} The probability of a stellar-driven instability ({\it blue}) occurring vs time. Orange shows the empirical \citet{abbot23} fit to the probability of an internally driven instability vs time.}
\label{fig:InstabTimes}
\end{figure}

Figure \ref{fig:InstabTimes}A plots the distribution of instability times (the time at which the first planet is lost) for our {\it strongpass} and {\it strongpass\_withstars} simulations. This distribution is roughly consistent with a uniform probability over time. The median instability time is 3.08 Gyrs, or $\sim$62\% of the total system integration. In Figure \ref{fig:InstabTimes}B, we compare the probability of a stellar-driven instability over time (inferred from Panel A's systems) against the probability of an internally driven instability (estimated with the \citet{abbot23} fit to prior ensembles of simulations). We see that for most of the next 5 Gyrs, the probability of a stellar-driven instability is higher than an internally driven one. The probability of an internally driven instability is several times smaller than a stellar-driven one during the next 2 Gyrs, and it does not exceed the stellar-driven probability until 4--4.5 Gyrs from now, when the Sun's main sequence lifetime is nearly over. 

\begin{figure}
\centering
\includegraphics[scale=0.42]{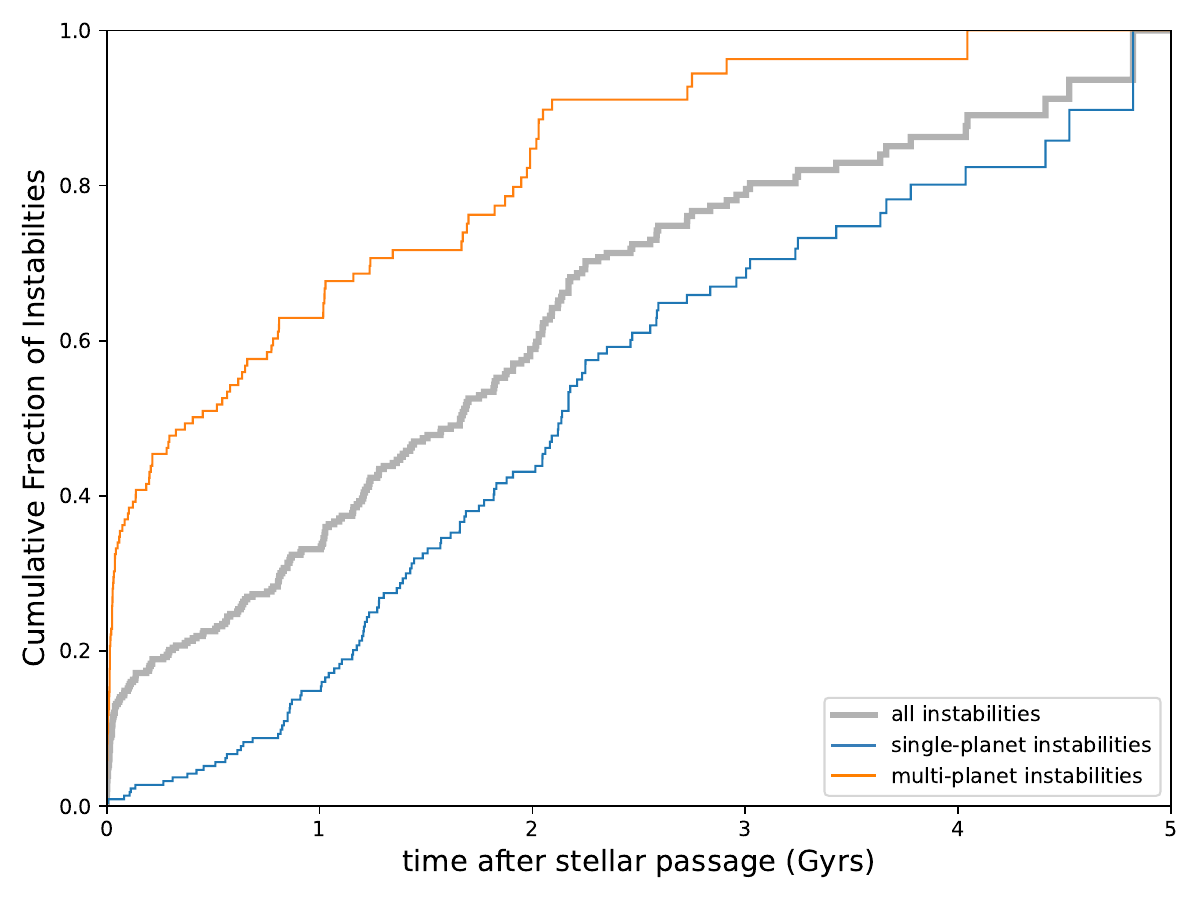}
\caption{The distribution of the time delay between a powerful stellar passage and the first planet loss among our 225 unstable {\it strongpass} and {\it strongpass\_withstars} systems. Gray shows the distribution for all unstable systems. Blue shows the distribution for systems that lose just one planet, and orange shows the distribution for systems that lose more than one planet.}
\label{fig:InstabDelays}
\end{figure}

It should also be noted that the median stellar-driven instability time in Figure \ref{fig:InstabTimes}A is not right at 2.5 Gyrs even though our powerful stellar encounters are randomly distributed over the next 5 Gyrs. This implies that there is, in fact, somewhat of a delay between the stellar passage and the subsequent planetary instability that it triggers. In Figure \ref{fig:InstabDelays}, we plot the distribution of delays between powerful stellar passages and the initiation of instabilities (first planet loss) among our {\it strongpass} and {\it strongpass\_withstars} systems. We see that there is actually a huge range in the lengths of time that elapse between the stellar passage and the instability. While the median is $\sim$1.7 Gyrs, instabilities can occur within a few Myrs and as late as 4--5 Gyrs after stellar passages. (We note that the delay values in Figure \ref{fig:InstabDelays}'s CDF are weighted according to the fraction of {\it strongpass} and {\it strongpass\_withstars} systems that are integrated longer than a given delay value, since these systems are integrated for random times between 0 and 5 Gyrs after being exposed to their stellar passage.) This large spread in delay times means complete assessments of the destabilizing effects of field star passages require at least $\sim$Gyr integration times, and employing shorter integration lengths will lead to underestimates of the instability probability \citep{laughadams00, ray24}. 

It is also apparent in Figure \ref{fig:InstabDelays} that the delays for single-planet and multi-planet instabilities are distributed quite differently. For single-planet instabilities, only $\sim$15\% occur within the first Gyr after the stellar passage, and the median delay is over 2.1 Gyrs. In contrast, among multi-planet instabilities, over 1/3 become unstable during the first 100 Myrs after the stellar passage, and the median delay is just over 400 Myrs. Nonetheless, multi-planet instabilities can also take substantial time to develop, and $\sim$1/3 begin 1 Gyr or more after the stellar passage has occurred. 

\begin{figure}
\centering
\includegraphics[scale=0.42]{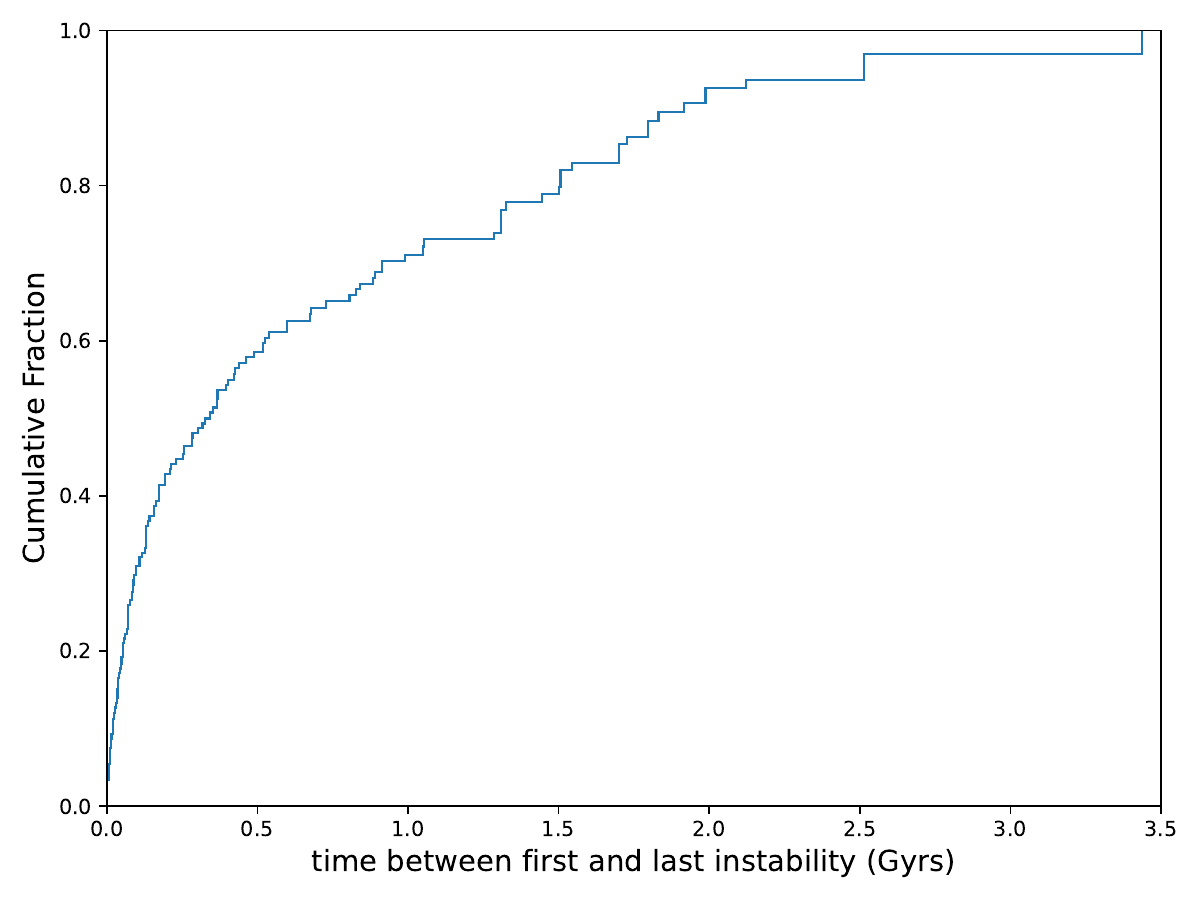}
\caption{Among our 111 {\it strongpass} and {\it strongpass\_withstars} systems that lose more than one planet, the distribution of times between system's first planet loss and last planet loss is shown.}
\label{fig:InstabDuration}
\end{figure}

In the case of multi-planet instabilities, it is possible for the solar system to enter a ``dynamically active'' state for a significant period of time over which planets are lost. In Figure \ref{fig:InstabDuration}, we plot the distribution of times between when systems lose their first planet and their last planet for all of our {\it strongpass} and {\it strongpass\_withstars} systems that feature multi-planet instabilities. (Again, this distribution is weighted according to the fraction of systems that feature total integrations longer than the period between a stellar encounter and the time of last planet loss.) In this distribution, we see that half of all multi-planet instabilities have multiple planetary losses over periods of 350 Myrs or longer, and approximately 30\% have at least 1 Gyr elapse between the first and last planetary loss. Thus, stellar passages can transform systems into a dynamically active state that lasts for significant fractions of the solar system age. We should note that the timescales in Figure \ref{fig:InstabDuration} represent a lower estimate of systems' dynamically active phases, as planet pairs typically begin exchanging energy via close encounters well before any planetary losses occur. We can use our simulations to measure the first time that two planets come within 1 Hill radius of one another relative to the time of first planetary loss. The median value for this time offset is just over 30 Myrs (for both single- or multi-planet instabilities), indicating that the typical lengths of the ``dynamically active'' phases of systems are approximately equal with either measure.

\subsection{Tail Cases For Earth}

Finally, we focus on the potential fates of Earth in our {\it strongpass} and {\it strongpass\_withstars} simulations. As noted in Figure \ref{fig:RelativeInstabRates}, Earth is $\sim$10 times less likely to be lost than Mercury, the most unstable planet. This works out to a loss probability of just $0.058^{+0.028}_{-0.020}$\% over the next 5 Gyrs. However, as previously discussed, we also find that the other inner solar system planets are significantly less stable that previously thought, and Earth collision can be the source of loss of some of those planets. We find that 64 of our 2000 {\it strongpass} and {\it strongpass\_withstars} systems feature a collision between Earth and another planet (typically Venus). Thus, there is a $0.22^{+0.05}_{-0.04}$\% chance that Earth will be ejected or collide with another planet over the next 5 Gyrs. This probability, while small, is $\sim$1--3 orders of magnitude higher than previous estimates \citep{ray24, laughadams00}. The reasons for the higher probability are primarily our longer integration lengths, consideration of stellar encounters that are more distant but still retain a large impulse gradient, and the inclusion of all four giant planets in all of our systems. 

\begin{figure}
\centering
\includegraphics[scale=0.42]{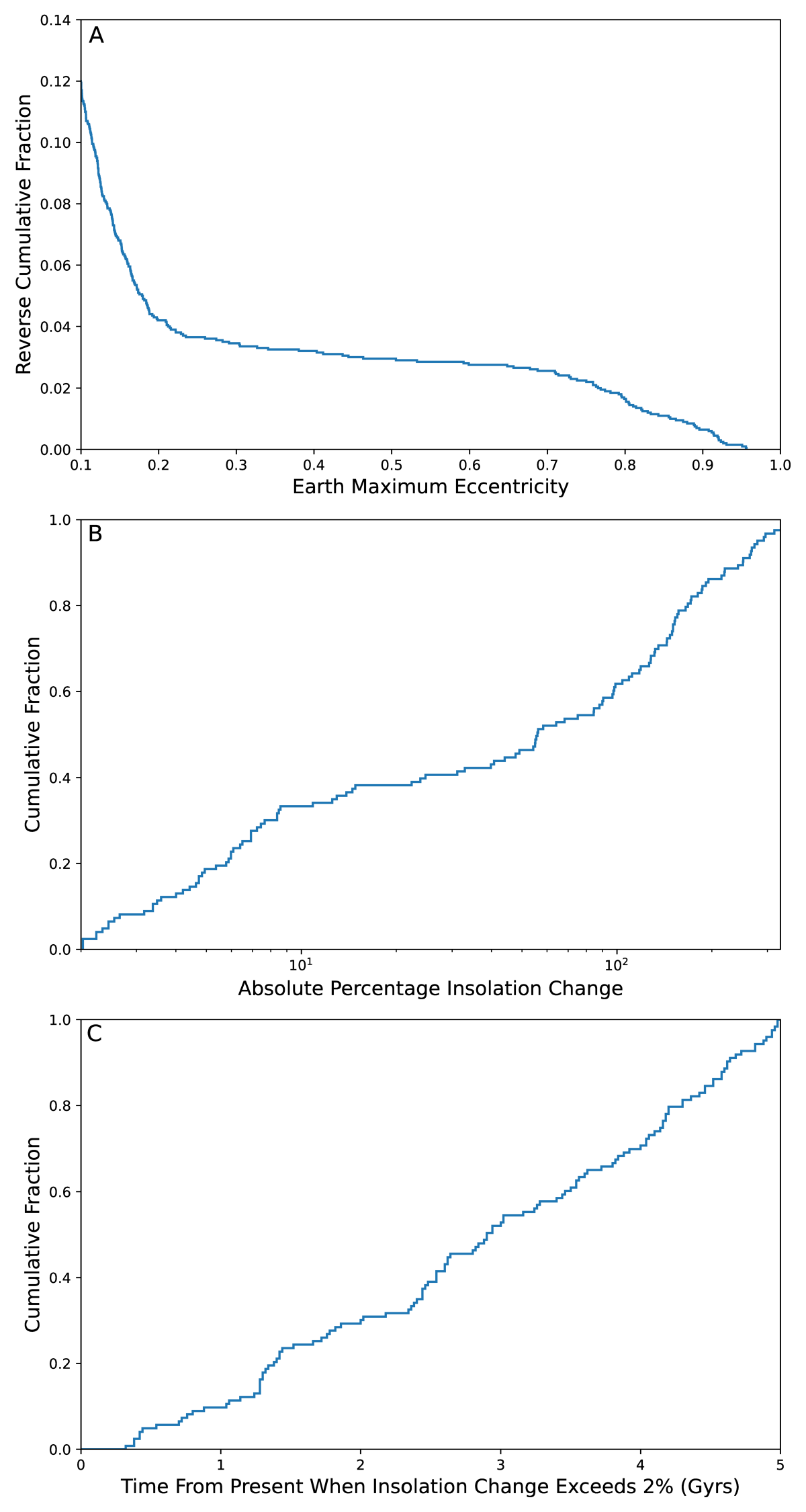}
\caption{{\bf A:} The distribution of the maximum eccentricity recorded for Earth for each system among our 2000 {\it withstars} and {\it strongpass\_withstars} systems. {\bf B:} For systems where Earth's orbital changes result in a 2\% or greater change in orbit-averaged insolation, the distribution of absolute insolation change is plotted. {\bf C:} For the same systems from panel B, the distribution of times when orbit-averaged insolation first changes by 2\% or more is plotted.}
\label{fig:EarthFlux}
\end{figure}

Of course, the Earth does not need to be ejected or collide with a planet for our planet's habitability to be significantly altered. In Figure \ref{fig:EarthFlux}A, we plot the distribution of maximum eccentricity that Earth attains in each of our 2000 {\it strongpass} and {\it strongpass\_withstars} systems. We find that in $4^{+1}_{-0.8}$\% of our systems (equivalent to an overall probability of $0.2^{+0.05}_{-0.04}$\%), Earth attains an eccentricity of 0.2 or higher. Such values are virtually never seen in integrations of the solar system in isolation \citep{lask08}. Moreover, our probability of 0.2\% is $\sim$200--300 times larger than prior estimates of such eccentricity excursions \citep{laughadams00}. An eccentricity of 0.2 would result in a $\sim$2\% increase in orbit-averaged insolation. This is $\sim$10 times larger than the eccentricity-induced insolation fluctuations expected over the next Gyr when considering the solar system in isolation \citep{lask08}. In addition, insolation can be altered through changes to Earth's semimajor axis as well, which are possible when other inner solar system planets are becoming unstable. In total, we find 123 of 2000 {\it strongpass} and {\it strongpass\_withstars} systems experience an insolation change greater than 2\%. This equates to an effective probability of $0.31\pm0.05$\% over the next 5 Gyrs. 

In Figure \ref{fig:EarthFlux}B, we plot the distribution of Earth insolation changes for {\it strongpass} and {\it strongpass\_withstars} systems that experience 2\% changes or greater. We see that most changes are actually much larger than 2\%. In fact, the median change for these systems is 56\%, which would certainly have dramatic effects on the terrestrial climate \citep{willpoll02, dress10, bol16} that could negatively impact planetary habitability \citep{pal20}. In Figure \ref{fig:EarthFlux}C, we plot the distribution of times at which systems' absolute insolation changes first exceed 2\%. We see that the median time is $\sim$2.9 Gyrs from now. Only 36 of our 2000 systems experience these changes within the first 2 Gyrs, after which point Earth is expected to leave the Sun's habitable zone \citep{lec13, wolftoon14, gra24}. Thus, factoring in the probabilities of the powerful stellar encounters employed in these simulation sets, we conclude that there is a $0.09^{+0.03}_{-0.02}$\% probability, or a $\sim$1 in 1000 chance, that Earth's total time as a habitable planet will be altered by a field star passage.

\section{Summary \& Conclusions}\label{sec:con}

To study the influence of passing field stars on the dynamical future of the solar system, we perform 5-Gyr simulations of the eight planets and Pluto beginning from orbital realizations consistent with the modern solar system. Our simulations indicate that stellar passage effects typically scale with the impulse gradient of the most powerful stellar encounter that the solar system experiences, and they alter the future evolution of the solar system in a number of significant ways. First, the solar system's potential for instability is significantly enhanced. For instance, in the presence of field stars, Pluto has a $3.9^{+1.3}_{-1.0}$\% chance of becoming ejected and another 1-2\% chance of being dislodged from its 3:2 resonance with Neptune (see Figure \ref{fig:plutoloss}).  Meanwhile, in the absence of passing stars Pluto appears to have a 0\% chance of being lost on 5-Gyr timescales \citep{itotan02, malito22}. 

This increase in instability probability extends to the eight planets as well. Prior work estimates that the eight planets' internal dynamics have a $\sim$0.8--1\% chance of driving Mercury into a collision with the Sun or Venus over the next 5 Gyrs \citep{abbot23}. Our work shows that stellar passages can also destabilize Mercury with a nearly comparable probability of $0.56\pm0.08$\%. Moreover, other planets are also susceptible to stellar-driven instabilities. Mars has a $0.28\pm0.05$\% chance of becoming unstable over the next 5 Gyrs, and Venus also has a $0.18^{+0.05}_{0.03}$\% chance of instability. For the ice giants and Earth, the instability probability is $\sim$0.05\% but still an order of magnitude or higher than prior estimates \citep{laughadams00, ray24}. Saturn's instability probability falls below 0.01\%, and Jupiter's is perhaps 1 in 40000 (see Figure \ref{fig:RelativeInstabRates}). Thus, destabilization via stellar passages is a major, underexplored dynamical pathway in the solar system's future evolution, and, relative to isolated solar system models, our planets' total instability probability increases by $\sim$50--80\%.

We also find that stellar-driven instabilities differ from previously studied internally driven instabilities in a number of important ways. The first is timing. Internally driven instabilities require several Gyrs of evolution before the instability probability rises rapidly \citep{abbot23}. In contrast, the probability of an instability-triggering stellar encounter is roughly uniformly distributed over the next 5 Gyrs, and our simulations show instability instances in the first Gyr. Thus, the probability of a stellar-driven instability is higher than the probability of internally driven one for the next 4--4.5 Gyrs, the large majority of the Sun's remaining main sequence lifetime (see Figure \ref{fig:InstabTimes}). 

In addition, we find that the nature of stellar-driven instabilities is more violent than internally driven ones. The loss of multiple planets in stellar-driven instabilities is common and occurs about 50\% of the time, whereas it appears quite rare for internally driven instabilities. The two planets most often lost are Mercury and Mars, but Venus also has a significant loss rate (see Figure \ref{fig:ExampInstab}). While Earth's instability rate is lower, its orbit can be significantly altered via scattering events or collisions with the lost terrestrial planets. We find a $0.31\pm0.05$\% chance that Earth's orbit will be modified to change its orbit-averaged insolation by 2\% or more (often much more) over the next 5 Gyrs, although $\sim$70\% of the time this change happens 2 Gyr or more from now, at which point Earth will likely no longer be habitable anyways (see Figure \ref{fig:EarthFlux}). Nonetheless, this probability of Earth orbital change is hundreds of times larger than prior estimates \citep{laughadams00}. This is mostly due to our longer integration times, inclusion of all planets, and consideration of a broad variety of stellar encounters. 

While the giant planets are very unlikely to become unstable, their orbits are not insulated from the influence of field stars. Our simulations show that stellar passages can drive a diffusion in the giant planets' secular frequencies and their corresponding amplitudes that can be a factor of a few to a factor of $\sim$100 greater than the diffusion solely due to the solar system's internal dynamics (see Figure \ref{fig:gmodediffuse}). In particular, the secular eigenfrequency associated with Uranus' eccentricity is the most sensitive to stellar passages. Even the most lightly perturbed systems feature twice as much chaotic diffusion as isolated systems, and our median system has an order of magnitude greater diffusion than the solar system in isolation. The reason for this is Uranus' proximity to the 2:1 MMR with Neptune. Field star passages drive small shifts in the semimajor axes of both planets, leading to an outsized effect on Uranus' precession frequency as it is pushed closer to or further from the resonance (see Figure \ref{fig:UResDriver}). We also find that the 2:1 resonance itself is a large source of solar system instability. Although the detailed mechanism is not clear, if a stellar passage happens to push Uranus and Neptune to within 1\% of their 2:1 MMR, the probability of a planetary instability rises from $\sim$1\% to over 50\% (see Figure \ref{fig:UNRatInstab}). 

In summary, passing stars can alter the stability of the planets and Pluto as well as the secular architecture of the giant planets over the next 5 Gyrs. Their significance on the solar system's dynamical future largely depends on the strength of the most powerful stellar passage over this time span, which is uncertain by orders of magnitude. This uncertainty in the Sun's future powerful stellar encounters means that the spectrum of future secular evolution and planetary instabilities is  broader than that implied by isolated models of solar system evolution. 

\section{Acknowledgements}

NAK's contributions to this work were supported from NASA Solar System Workings grant 80NSSC24K1874. SNR acknowledges funding from the Programme Nationale de Planetologie (PNP) of the INSU (CNRS), and in the framework of the Investments for the Future programme IdEx, Universite de Bordeaux/RRI ORIGINS. This research was done using services provided by the OSG Consortium \citep{osg07, osg09, link1, link2}, which is supported by the National Science Foundation awards \#2030508 and \#2323298. Finally, we thank both anonymous reviewers, whose comments greatly improved this work. \clearpage

\bibliographystyle{apj}
\bibliography{SSStability}

\end{document}